\documentclass[12pt,a4paper,aps,preprint,superscriptaddress,nofootinbib]{revtex4-1}
\usepackage[utf8]{inputenc}
\usepackage{graphicx}
\usepackage{amssymb}
\usepackage{textcomp}
\usepackage{amsmath}
\usepackage{tabularx}
\usepackage{bm}
\usepackage{times}
\usepackage{color}
\usepackage{slashed}
\usepackage{multirow}
\usepackage{verbatim}
\usepackage{cancel}
\usepackage{subfigure}
\usepackage[normalem]{ulem}
\usepackage{xcolor}
\usepackage{float}

\usepackage[colorlinks=true, pdfstartview=FitV, linkcolor=blue, citecolor=blue, urlcolor=blue]{hyperref}

\allowdisplaybreaks[4]

\def\lsim{\mathrel{\raise.3ex\hbox{$<$\kern-.75em\lower1ex\hbox{$\sim$}}}}
\def\gsim{\mathrel{\raise.3ex\hbox{$>$\kern-.75em\lower1ex\hbox{$\sim$}}}}

\definecolor{orange}{rgb}{1,0.5,0}

\begin{document}

\title{Probing eV-scale sterile neutrinos with neutrinoless double-beta decay and beta decay}

\author{Ruisha Yang}
\email{yangrsh5@mail2.sysu.edu.cn}
\affiliation{
	School of Physics, Sun Yat-Sen University, Guangzhou 510275, China
}

\author{Jiajun Liao}
\email{liaojiajun@mail.sysu.edu.cn}
\affiliation{
	School of Physics, Sun Yat-Sen University, Guangzhou 510275, China
}

\begin{abstract}
An eV-scale sterile Majorana neutrino can suppress neutrinoless double-beta decay through destructive interference with the active-neutrino contribution, weakening the resulting constraints on sterile-neutrino parameters. We investigate how complementary beta-decay information restricts this cancellation in the 3+1 framework. Using updated oscillation parameters that include the first JUNO results, we determine the allowed sterile-neutrino mass and mixing regions for both normal and inverted ordering. We show that an upper limit on the effective beta-decay mass $m_\beta$ of the three light neutrinos strengthens the constraints from neutrinoless double-beta decay alone. The beta-decay bound restricts the light-neutrino masses and hence the cancellation between the active and sterile contributions to $m_{\beta\beta}$. We derive analytic expressions for the resulting boundaries and use complex-plane geometry to explain the allowed values of the sterile Majorana phase $\alpha_4$ for a finite $|m_{\beta\beta}|$ limit. For the mixing considered, normal ordering allows a broad range of $\alpha_4$ at small sterile masses, whereas inverted ordering favors values near $\pi$. The allowed phase range narrows as the sterile mass approaches its upper bound or the limit on $|m_{\beta\beta}|$ becomes more stringent.
\end{abstract}

\maketitle

\tableofcontents

\newpage

\section{Introduction}
\label{sec:Intro}

The observation of neutrino oscillations has established nonzero neutrino masses and lepton flavor mixing, providing compelling evidence of physics beyond the Standard Model (SM)~\cite{Gonzalez-Garcia:2002bkq,SNO:2002tuh,K2K:2002icj,KamLAND:2004mhv}.
However, oscillation experiments are primarily sensitive to the mass-squared differences and mixing parameters, leaving fundamental questions about the absolute neutrino mass scale and the Dirac or Majorana nature of neutrinos unresolved. Neutrinoless double-beta decay ($0\nu\beta\beta$) offers a direct test of total lepton-number violation and provides an important probe of the Majorana nature of neutrino mass~\cite{Jones:2021cga, Bilenky:2014uka,Huang:2023tjy,Dery:2024lem}.
Throughout this work, we assume that light-Majorana-neutrino exchange dominates the $0\nu\beta\beta$ amplitude. The decay rate is then governed by the effective Majorana mass $m_{\beta\beta}$~\cite{Benato:2015via}, which depends on the absolute neutrino mass scale, the mass ordering, and the Majorana phases~\cite{Vergados:2016hso}. The inverse half-life of $0 \nu \beta\beta$ is
\begin{equation}
	\left( T^{0\nu}_{1/2} \right)^{-1}=G^{0\nu}g_{A}^4|M^{0\nu}|^2\frac{\left | m_{\beta\beta} \right | ^2}{m_{e}^2} \,,
\end{equation}
where $G^{0\nu}$ is the phase-space factor~\cite{Kotila:2012zza,Mirea:2014dza}, $g_{A}$ is the axial-vector coupling constant~\cite{Bilenky:2014uka}, $\left |M^{0\nu}  \right |$ is the nuclear matrix element (NME)~\cite{Agostini:2022zub}, and $m_{e}$ is the electron mass. Converting a half-life limit into a bound on $|m_{\beta\beta}|$ entails substantial theoretical uncertainty, primarily from the NME. NMEs are calculated using several nuclear-structure approaches, including the quasiparticle random-phase approximation (QRPA)~\cite{Faessler:2007hu,Engel:2014pha}, the nuclear shell model (NSM)~\cite{Kota:2025dzk}, the interacting boson model (IBM-2)~\cite{Barea:2013bz}, and energy-density-functional methods (EDF)~\cite{Ding:2024obt}. The predictions from different nuclear structure models vary significantly, leading to different inferred bounds on $m_{\beta\beta}$.
Current $0\nu\beta\beta$ searches are approaching the effective-mass range associated with inverted ordering in the standard three-neutrino framework. Experiments such as KamLAND-Zen~\cite{KamLAND-Zen:2022tow} and CUORE~\cite{CUORE:2024ikf} have set stringent lower bounds on the half-life, but no conclusive signal has been observed.
These null results motivate tests of both the standard three-neutrino framework and its extensions.

Direct beta-decay measurements provide information complementary to $0\nu\beta\beta$ searches~\cite{Formaggio:2021nfz}. The kinematic effective mass $m_{\beta}$, probed through tritium $\beta$ decay, provides a phase-independent constraint on the absolute neutrino-mass scale.
Unlike $m_{\beta\beta}$, $m_{\beta}$ does not depend on the Majorana phases and can be inferred from a fit to the endpoint region of the $\beta$-decay spectrum~\cite{Kleesiek:2018mel}. The KATRIN experiment has set an upper limit of $m_{\beta} < 0.45\,\mathrm{eV}$ at 90\%\ C.L.~\cite{KATRIN:2024cdt}, while Project~8 aims to reach a sensitivity of approximately $0.04\,\mathrm{eV}$~\cite{Project8:2022wqh}.


Cosmology also probes neutrino masses, but the inferred bounds depend on the cosmological model and data combination. Recent analyses of cosmic microwave background (CMB) and baryon acoustic oscillation (BAO) data show a tension with the minimum neutrino mass sum implied by oscillation measurements~\cite{Elbers:2025vlz,SPT-3G:2025bzu}. For example, the combined Planck, ACT, SPT-3G, and DESI DR2 analysis in flat $\Lambda$CDM gives $\Sigma\equiv\sum_{i=1}^{3}m_i<0.055\,\mathrm{eV}$ at 95\% C.L., with a prior $\Sigma\geq0$~\cite{SPT-3G:2025bzu}. Oscillation measurements instead require $\Sigma\gtrsim0.059\,\mathrm{eV}$ for normal ordering and $\Sigma\gtrsim0.10\,\mathrm{eV}$ for inverted ordering~\cite{Elbers:2025vlz}. The inferred limits vary with the CMB likelihood and can be relaxed by allowing an evolving dark-energy equation of state~\cite{Naredo-Tuero:2024sgf,Elbers:2025vlz}.

Over the past decades, neutrino oscillation measurements have progressively improved the precision of the three-neutrino mixing parameters. At the same time, several observations commonly referred to as ``short-baseline anomalies'' have motivated phenomenological tests of extensions to the standard three-neutrino framework.
The LSND experiment observed an excess of electron antineutrino events in the $\overline{\nu}_{\mu} \to \overline{\nu}_{e}$ channel~\cite{LSND:2001aii}, MiniBooNE observed an excess of low-energy electron-like events in both neutrino and antineutrino modes~\cite{MiniBooNE:2018esg}, and gallium source experiments including GALLEX~\cite{Kaether:2010ag}, SAGE~\cite{SAGE:2009eeu},  and BEST~\cite{Barinov:2021asz} measured a deficit of $\nu_{e}$ events.
These anomalies can be accommodated phenomenologically by oscillations driven by a new $\mathrm{eV^2}$-scale mass-squared difference~\cite{Gariazzo:2017fdh,Gariazzo:2019cym}.
A minimal phenomenological realization is to introduce one predominantly sterile state in addition to the three active neutrinos~\cite{Dentler:2018sju,Kopp:2013vaa,Goswami:2026qpl}. The sterile flavor is a singlet under the SM gauge group and has no SM weak coupling, although it may mix with the active flavors. The additional state modifies the relation between $m_{\beta\beta}$ and beta-decay observables, allowing lepton-number-violating and kinematic measurements to test complementary combinations of the sterile mass and mixing parameters~\cite{Meroni:2014tba,Giunti:2015iyr,Huang:2019qvq}. 

The relevant experimental information has recently improved along several complementary directions.
JUNO is designed to determine the solar oscillation parameters $\theta_{12}$ and $\Delta m^2_{21}$ with high precision~\cite{JUNO:2022mxj}.
The first JUNO reactor-neutrino results provide new precision measurements of these parameters~\cite{JUNO:2025gmd}.
The improved solar parameters reduce the parametric uncertainty in the active contribution to $m_{\beta\beta}$, particularly near cancellation regions.
At the same time, next-generation $0\nu\beta\beta$ experiments with larger exposures and lower backgrounds, such as nEXO~\cite{nEXO:2017nam} and CUPID~\cite{CUPID:2025faj}, aim to reach sensitivities near $m_{\beta\beta}\sim 10\,\mathrm{meV}$~\cite{Ge:2025cky}. 
Meanwhile, KATRIN~\cite{KATRIN:2024cdt, KATRIN:2021uub} and Project~8~\cite{Project8:2022wqh} provide direct, phase-independent kinematic information on the absolute neutrino-mass scale.
Combining these observables can reduce degeneracies among the absolute mass scale, the Majorana phases, and a possible sterile contribution.
This complementarity motivates us to investigate these constraints directly in the sterile-neutrino parameter plane.

The effects of sterile neutrinos on $0\nu\beta\beta$ have been studied extensively. Giunti and Zavanin systematically studied $m_{\beta\beta}$ in the three-neutrino and 3+1 scenarios, including its dependence on the lightest neutrino mass and its correlation with the beta-decay mass~\cite{Giunti:2015kza}. They showed that a sterile contribution can substantially alter the predicted effective-mass bands and remove the nonzero lower bound associated with inverted ordering in the three-neutrino framework.
The interplay among $0\nu\beta\beta$, tritium beta decay, and cosmology in the presence of an eV-scale sterile neutrino has also been investigated~\cite{Huang:2019qvq,Goswami:2005ng}. More recently, Jana, Puetter, and Smirnov derived a compact analytic relation that maps a limit on $m_{\beta\beta}$ onto the $(\sin^2 2\theta_{14},\Delta m_{41}^2)$ plane for representative light-neutrino spectra~\cite{Jana:2024xmc}. They examined how the resulting boundaries depend on the light-neutrino mass ordering and absolute mass scale, and compared the $0\nu\beta\beta$ constraints with other experimental indications and exclusion limits.

Here we examine how a light-sector beta-decay mass bound restricts the sterile-neutrino regions compatible with a prospective $0\nu\beta\beta$ limit, using the 2025 oscillation inputs that include the first JUNO results.
We also assess the effect of the updated oscillation inputs on the effective-mass predictions and the boundaries in the $(\sin^2 2\theta_{14},\Delta m_{41}^2)$ plane.
Rather than restricting the analysis to a few representative light-neutrino spectra, we vary the lightest neutrino mass continuously and the standard oscillation parameters within their updated $3\sigma$ intervals. 
We compare the allowed regions obtained from $0\nu\beta\beta$ alone with those obtained after including an upper bound on the light-sector kinematic mass.
Since recent analyses of CMB and BAO data show a tension with the minimum neutrino mass sum implied by oscillation measurements~\cite{Elbers:2025vlz,SPT-3G:2025bzu}, we only consider terrestrial experiments, and use oscillation data together with prospective $0\nu\beta\beta$ and beta-decay limits to constrain the sterile-neutrino parameters.

The sterile Majorana phase $\alpha_4$ is central to this interplay because it controls the interference between the sterile and active contributions to $m_{\beta\beta}$.
We derive a geometric condition for the allowed values of $\alpha_4$ and study their dependence on $m_4$ and the upper limit on $|m_{\beta\beta}|$. This condition determines the allowed phase intervals for a finite upper limit on $|m_{\beta\beta}|$ and explains their dependence on the sterile mass and mass ordering.
This geometric analysis explains the origin and boundaries of the phase regions in normal and inverted ordering for a finite effective-mass bound, rather than only describing exact cancellation.

The paper is organized as follows. In Sec.~\ref{sec:framework}, we discuss the three-neutrino and 3+1 frameworks and define the beta-decay quantities used in the analysis. In Sec.~\ref{sec:analysis}, we present our main results. In Sec.~\ref{sec:fixedparameters}, we examine the effective-mass correlations for fixed sterile-neutrino parameters. In Sec.~\ref{sec:Restrictsterileparameter}, we map prospective mass limits to compatibility regions in the sterile-neutrino parameter plane for both mass orderings. In Sec.~\ref{sec:alpha4}, we study the mass and sensitivity dependence of the compatible sterile Majorana phase $\alpha_4$ and interpret the resulting regions geometrically.
We summarize our conclusions in Sec.~\ref{sec:conclusion}.

\section{Framework}
\label{sec:framework}

In this section, we first define the mass and mixing conventions in the three-neutrino and 3+1 frameworks, then specify the beta-decay observables used below.

\subsection{Three-neutrino mixing}
\label{threeneumix}

In the standard three-neutrino framework, the flavor eigenstates $\nu_\alpha$ $(\alpha=e,\mu,\tau)$ are related to the mass eigenstates $\nu_i$ $(i=1,2,3)$ by the Pontecorvo--Maki--Nakagawa--Sakata (PMNS) mixing matrix, 
\begin{equation}
\nu_\alpha=\sum_{i=1}^{3}U_{\alpha i}\nu_i \,.
\label{eq:flavor_mass_relation}
\end{equation}

For three Majorana neutrinos, the PMNS matrix can be written as~\cite{Giganti:2017fhf,Qian:2015waa}
\begin{equation}
U =
\begin{pmatrix}
c_{12}c_{13}
&
s_{12}c_{13}
&
s_{13}e^{-i\delta_{\rm CP}}
\\[2mm]
-s_{12}c_{23}-c_{12}s_{23}s_{13}e^{i\delta_{\rm CP}}
&
c_{12}c_{23}-s_{12}s_{23}s_{13}e^{i\delta_{\rm CP}}
&
s_{23}c_{13}
\\[2mm]
s_{12}s_{23}-c_{12}c_{23}s_{13}e^{i\delta_{\rm CP}}
&
-c_{12}s_{23}-s_{12}c_{23}s_{13}e^{i\delta_{\rm CP}}
&
c_{23}c_{13}
\end{pmatrix}
P_{\rm M}\,,
\label{eq:PMNS_matrix}
\end{equation}
where
$
P_{\rm M}
=
\operatorname{diag}
\left(
1,\,
e^{i\alpha_{2}/2},\,
e^{i\alpha_{3}/2}
\right)
$.
Here, $s_{ij}\equiv\sin\theta_{ij}$ and
$c_{ij}\equiv\cos\theta_{ij}$, $\delta_{\rm CP}$ is the Dirac CP-violating phase, and $\alpha_{2}$ and $\alpha_{3}$ are the two Majorana phases.
Only two independent phase combinations enter $m_{\beta\beta}$, and the Dirac phase can be absorbed into one of them. We denote the resulting effective Majorana-phase combinations by $\alpha_2$ and $\alpha_3$ below, so no separate input for $\delta_{\rm CP}$ is required.

We use the mass-squared-difference convention adopted in the oscillation analyses of Refs.~\cite{Giunti:2015kza,Capozzi:2025wyn}. The solar and atmospheric mass-squared differences are defined as
\begin{equation}
\label{eq:msquare}
    \begin{split}
    &\delta m^2=m_{2}^2-m_{1}^2\,,\\
    &\Delta m^2=m_{3}^2-\frac{1}{2} (m_{1}^2+m_{2}^2)\,.
    \end{split}
\end{equation}

With this convention, $\Delta m^2$ is positive for normal ordering (NO), for which $m_1<m_2<m_3$, and negative for inverted ordering
(IO), for which $m_3<m_1<m_2$. 
Eq.~(\ref{eq:msquare}) determines the three masses in terms of the lightest mass and the two measured mass-squared differences.

Assuming that $0\nu\beta\beta$ is dominated by the exchange of light Majorana neutrinos, the effective Majorana mass is~\cite{Giunti:2015kza}
\begin{equation}
\label{eq:effectivemass}
    \left |m_{\beta\beta} \right |=\left |\mu_{1} + \mu_{2}e^{i\alpha_{2}}+\mu_{3}e^{i\alpha_{3}} \right | \,,
\end{equation}
where 
$
    \mu_{k}=m_{k}\left |U_{ek}\right |^2
$ denotes the magnitude of the $k$th mass-eigenstate contribution.
The magnitudes of the relevant electron-row elements are
\begin{equation} 
    \left | U_{e_{1}} \right |=\cos\theta_{12}\cos\theta_{13}\,,
    \left | U_{e_{2}} \right |=\sin\theta_{12}\cos\theta_{13}\,,
    \left | U_{e_{3}} \right |=\sin\theta_{13}\,.
\end{equation}


For fixed masses and mixing angles, varying the Majorana phases determines the allowed interval of $|m_{\beta\beta}|$.
The maximum occurs when all three contributions are aligned in the complex plane~\cite{Giunti:2015kza,Vissani:1999tu},
\begin{equation}
     |m_{\beta\beta}|_{\max}
    =
    \sum_{i=1}^{3}\mu_i \,,
\end{equation}
whereas maximal destructive interference gives the minimum
\begin{equation}
\label{eq:mmin}
    |m_{\beta\beta}|_{\min}
    =
    \max
    \left[
    2\mu_{\max}
    -
    \sum_{i=1}^{3}\mu_i,
    0
    \right],
    \qquad
    \mu_{\max}
    =
    \max\{\mu_1,\mu_2,\mu_3\}\,.
\end{equation} 

The complementary kinematic mass probed by tritium beta decay is independent of the Majorana phases.
When the three mass eigenstates are unresolved experimentally, their combined effect near the endpoint is characterized by
\begin{equation}
\label{eq:kinematicmass}
    m_{\beta}=\sqrt{m_{1}^2\left | U_{e_{1}} \right |^2+m_{2}^2\left | U_{e_{2}} \right |^2+m_{3}^2\left | U_{e_{3}} \right |^2}\,.
\end{equation}

The two observables probe complementary combinations of masses and mixing parameters, with $m_{\beta\beta}$ additionally sensitive to the Majorana phases.
Their predicted ranges depend on the oscillation inputs $\theta_{12}$, $\theta_{13}$, $\delta m^2$ and $\Delta m^2$.


\subsection{The 3+1 mixing framework}
\label{sec:the31model}

The three-neutrino framework can be minimally extended by adding a sterile flavor state. The additional mass eigenstate introduces an oscillation frequency associated with a mass-squared difference larger than the solar and atmospheric splittings~\cite{Giunti:2013aea,Kopp:2013vaa,Meroni:2014tba}. We take this additional state to be predominantly sterile.
Although the sterile flavor has no SM weak interaction, it can mix with the active flavors. We denote the predominantly sterile mass eigenstate by $\nu_4$. We define the additional mass-squared difference as $\Delta m_{41}^2\equiv m_4^2-m_1^2$ in both orderings, so that $m_4=\sqrt{m_1^2+\Delta m_{41}^2}$.

In the 3+1 framework, the lepton mixing matrix becomes a $4\times4$ unitary matrix. A convenient parameterization of the $4\times4$ mixing matrix is
\begin{equation}
    U
    =
    R_{34}\widetilde R_{24}\widetilde R_{14}
    R_{23}\widetilde R_{13}R_{12}P_M \,,
\end{equation}
where $R_{ij}$ and $\widetilde R_{ij}$ denote real and complex rotations, respectively, and
\begin{equation}
    P_M
    =
    {\rm diag}
    \left(
    1,
    e^{i\alpha_2/2},
    e^{i\alpha_3/2},
    e^{i\alpha_4/2}
    \right)\,,
\end{equation}
contains the three Majorana phases.
The complex rotations $\widetilde R_{ij}$ contain the corresponding Dirac-type phases. As in the three-neutrino case, we absorb the Dirac phases entering the electron row into the three phase combinations relevant to $m_{\beta\beta}$ and denote them by $\alpha_2$, $\alpha_3$, and $\alpha_4$. The parameters associated exclusively with the $\mu$- and $\tau$-sterile sectors do not enter the observables studied below.
The first-row elements describing electron-flavor mixing can be parameterized as
\begin{equation}
\label{eq:PMNS}
    \begin{split}
        &\left | U_{e_{1}} \right | =\cos\theta_{14}\cos\theta_{13}\cos\theta_{12}\,,\\
        &\left | U_{e_{2}} \right | =\cos\theta_{14}\cos\theta_{13}\sin\theta_{12}\,,\\
        &\left | U_{e_{3}} \right | =\cos\theta_{14}\sin\theta_{13}\,,\\
        &\left | U_{e_{4}} \right |=\sin\theta_{14}\,.
    \end{split}
\end{equation}

Here, the mixing angle $\theta_{14}$ characterizes the mixing of the electron flavor with the fourth mass eigenstate. 
In the 3+1 framework, the effective Majorana mass for $0\nu\beta\beta$ decay is given by~\cite{Giunti:2015kza}
\begin{equation}
\label{eq:sterilembetabeta}
    \left |m_{\beta\beta}\right |=\left |\mu_{1} + \mu_{2}e^{i\alpha_{2}}+\mu_{3}e^{i\alpha_{3}}+\mu_{4}e^{i\alpha_{4}} \right |\,,
\end{equation}
where $
    \mu_k
    \equiv
    m_k|U_{ek}|^2
$.
The fourth state adds a contribution with magnitude $\mu_4=m_4\sin^2\theta_{14}$ and phase $\alpha_4$. Its interference with the active contribution changes the cancellation conditions for $m_{\beta\beta}$.

Similar to the three-neutrino case, the maximum effective mass is obtained when the four contributions are aligned,
\begin{equation}
    |m_{\beta\beta}|_{\max}
    =
    \sum_{i=1}^{4}\mu_i \,,
\end{equation}
whereas the minimum over the full Majorana-phase space is
\begin{equation}
\label{eq:mbb_min}
    |m_{\beta\beta}|_{\min}
    =
    \max
    \left[
    2\mu_{\max}
    -
    \sum_{i=1}^{4}\mu_i,
    0
    \right],
    \qquad
    \mu_{\max}
    =
    \max\{\mu_1,\mu_2,\mu_3,\mu_4\}\,.
\end{equation}

Complete cancellation is possible when no single contribution exceeds the sum of the remaining three. Otherwise, the minimum effective mass is given
by the excess of the largest contribution over the other three.


\subsection{Beta-decay effective mass in the 3+1 framework}
\label{sec:betakink}

Direct beta-decay experiments probe the absolute neutrino-mass scale through the distortion of the electron spectrum near the endpoint. In the standard three-neutrino framework, when the three light mass eigenstates cannot be resolved individually, their combined effect can be described by the effective kinematic mass
\begin{equation}
m_{\beta}^{2}
=
\sum_{i=1}^{3}
|\widetilde U_{ei}|^{2}m_i^{2}
=
c_{13}^{2}c_{12}^{2}m_{1}^{2}
+c_{13}^{2}s_{12}^{2}m_{2}^{2}
+s_{13}^{2}m_{3}^{2}\,,
\label{eq:mbeta3nu}
\end{equation}
where $\widetilde U$ denotes the standard three-neutrino mixing matrix. 
Unlike $m_{\beta\beta}$, the kinematic mass $m_\beta$ is independent of the Majorana phases and therefore provides complementary information on the light-neutrino mass scale.

In the $3+1$ framework, the contribution of the fourth mass eigenstate must be distinguished from that of the unresolved light states. For an eV-scale fourth state, the two contributions can produce distinct spectral distortions. Following Ref.~\cite{Giunti:2019fcj}, and within the effective-mass approximation for the light component, we write
\begin{equation}
R_\beta(E)
=
\left(1-|U_{e4}|^{2}\right)
R_\beta(E,m_\beta)
+
|U_{e4}|^{2}R_\beta(E,m_4)\,,
\label{eq:beta31spectrum}
\end{equation}
where $R_\beta(E,m_\beta)$ describes the combined contribution of the three light mass eigenstates, whereas $R_\beta(E,m_4)$ represents the contribution of the predominantly sterile state. 
The relative weight of the latter is determined by
$|U_{e4}|^{2}=s_{14}^{2}$.

Accordingly, the effective mass associated with the three unresolved light states is defined as~\cite{Giunti:2015kza,Farzan:2002zq,Giunti:2019fcj}
\begin{equation}
m_\beta^{2}
=
\sum_{i=1}^{3}
\frac{|U_{ei}|^{2}}
     {1-|U_{e4}|^{2}}
m_i^{2}\,.
\label{eq:mbeta31}
\end{equation}

Using the first-row elements of the $3+1$ mixing matrix in Eq.~(\ref{eq:PMNS}),
\begin{equation}
|U_{e1}|^{2}=c_{14}^{2}c_{13}^{2}c_{12}^{2},
\qquad
|U_{e2}|^{2}=c_{14}^{2}c_{13}^{2}s_{12}^{2},
\qquad
|U_{e3}|^{2}=c_{14}^{2}s_{13}^{2}\,,
\end{equation}
together with $1-|U_{e4}|^{2}=c_{14}^{2}$, Eq.~(\ref{eq:mbeta31}) becomes
\begin{equation}
m_{\beta}^{2}
=
c_{13}^{2}c_{12}^{2}m_{1}^{2}
+
c_{13}^{2}s_{12}^{2}m_{2}^{2}
+
s_{13}^{2}m_{3}^{2}\,.
\label{eq:mbeta_light}
\end{equation}

Thus, at fixed light-neutrino masses and standard mixing angles, $m_\beta$ has no explicit dependence on $\theta_{14}$ or $m_4$. Sterile mixing instead changes the relative weights of the two spectral components in Eq.~(\ref{eq:beta31spectrum}), while $m_4$ sets the threshold of the fourth-state contribution. These quantities can be varied as independent fit parameters, although they may be correlated in a spectral fit.

This separation is particularly relevant for an eV-scale sterile neutrino. 
Instead of simply contributing to the common light-neutrino endpoint mass, the fourth state can generate an additional spectral distortion whose position and magnitude depend primarily on $m_4$ and $|U_{e4}|^{2}$, respectively. 
Consequently, the quantity denoted by $m_\beta$ in the following analysis refers to the effective mass scale of the three light neutrinos, while the sterile-neutrino parameters are treated independently through $\theta_{14}$ and $\Delta m_{41}^{2}$. 

The distinction between the two quantities is also important for the joint analysis of
$0\nu\beta\beta$ and beta decay performed below.
The effective Majorana mass depends explicitly on the fourth-state contribution and on the additional Majorana phase $\alpha_4$, whereas $m_\beta$ in Eq.~(\ref{eq:mbeta_light}) constrains the absolute mass scale of the three light states independently of the Majorana phases. 
The two observables therefore provide complementary information: a bound on $m_\beta$ limits the light-neutrino contribution available for cancellation in $m_{\beta\beta}$, while the sterile contribution is controlled by $m_4$, $\theta_{14}$ and $\alpha_4$.

In the numerical analysis, we consider
\begin{equation}
\label{eq:mbetalimit}
m_\beta \leq 0.20~{\rm eV},
\qquad
m_\beta \leq 0.04~{\rm eV}\,,
\end{equation}
as representative light-sector mass benchmarks motivated by the design sensitivity of KATRIN~\cite{KATRIN:2021uub} and the target sensitivity of Project~8~\cite{Project8:2022wqh}, respectively.
These conditions are imposed on the light-sector quantity defined in Eq.~(\ref{eq:mbeta_light}).
We use these projected sensitivities as upper bounds on the normalized light-sector mass $m_\beta$, without performing a new $3+1$ beta-spectrum fit.

\section{Results}
\label{sec:analysis}

We now apply these definitions to the effective-mass correlations and sterile-neutrino parameter space. We first examine $|m_{\beta\beta}|$ and $m_\beta$ for a fixed sterile-neutrino benchmark.
We then determine the regions in the $(\sin^2 2\theta_{14},\Delta m_{41}^{2})$ plane compatible with the assumed effective-mass limits. In addition, we determine the compatible $\alpha_4$ intervals and use complex-plane geometry to explain their dependence on the sterile mass and the assumed $|m_{\beta\beta}|$ limit.

\subsection{Effective-mass correlations for fixed sterile parameters}
\label{sec:fixedparameters}
An eV-scale sterile neutrino can substantially modify the relation between $|m_{\beta\beta}|$ and the lightest mass, including by allowing complete cancellation in IO~\cite{Giunti:2015kza}. 
We first illustrate this interference effect at a fixed sterile-neutrino benchmark.
We choose $\sin^2 2\theta_{14}=0.0665$ and $\Delta m_{41}^{2}=4.0~\mathrm{eV}^2$, as marked in Fig.~\ref{fig:sterilespaceNH}. This benchmark point illustrates the interference pattern and is not excluded by the representative contours displayed in Fig.~\ref{fig:sterilespaceNH}, compiled from Ref.~\cite{KATRIN:2025lph}.

\begin{table}[b]
\centering
\caption{
$3\sigma$ ranges of the oscillation parameters used in the numerical analysis.
The JUNO-informed 2025 ranges of $\sin^2\theta_{12}$ and $\delta m^2$ are taken from the global fit including the SNO+ and first JUNO data~\cite{Capozzi:2025ovi}, while those of $\sin^2\theta_{13}$ and $|\Delta m^2|$ are taken from Ref.~\cite{Capozzi:2025wyn}.
The pre-JUNO 2024 ranges are taken from Ref.~\cite{Capozzi:2025wyn}.
}
\label{tab:parameter_ranges}

\begin{tabular}{lcccc}
\hline\hline
& \multicolumn{2}{c}{w/o JUNO 2024}
& \multicolumn{2}{c}{w/ JUNO 2025} \\
\cline{2-3}\cline{4-5}
Parameter & NO & IO & NO & IO \\
\hline

$\sin^2\theta_{12}$
&
$[0.264,\,0.345]$
&
$[0.264,\,0.345]$
&
$[0.2866,\,0.3303]$
&
$[0.2866,\,0.3303]$
\\

$\sin^2\theta_{13}$
&
$[0.0206,\,0.0238]$
&
$[0.0208,\,0.0241]$
&
$[0.0206,\,0.0238]$
&
$[0.0208,\,0.0241]$
\\

$\delta m^2/(10^{-5}\,\mathrm{eV}^2)$
&
$[6.93,\,7.93]$
&
$[6.93,\,7.93]$
&
$[7.21,\,7.78]$
&
$[7.21,\,7.78]$
\\

$|\Delta m^2|/(10^{-3}\,\mathrm{eV}^2)$
&
$[2.433,\,2.558]$
&
$[2.403,\,2.527]$
&
$[2.433,\,2.558]$
&
$[2.403,\,2.527]$
\\

\hline\hline
\end{tabular}
\end{table}

\begin{figure}[t]
	\centering
    \includegraphics[width=0.49\linewidth]{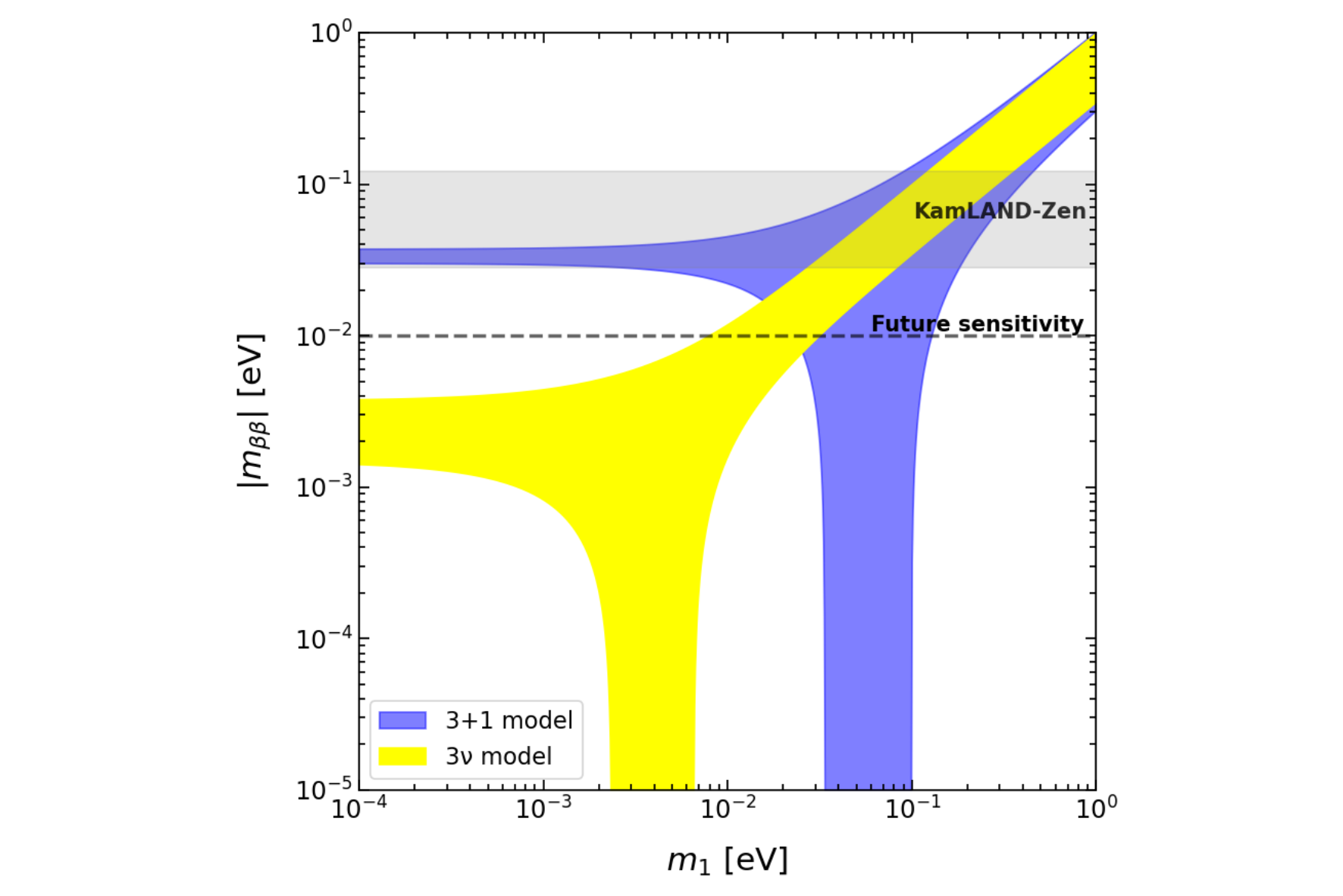}
	\hfill
	\includegraphics[width=0.49\linewidth]{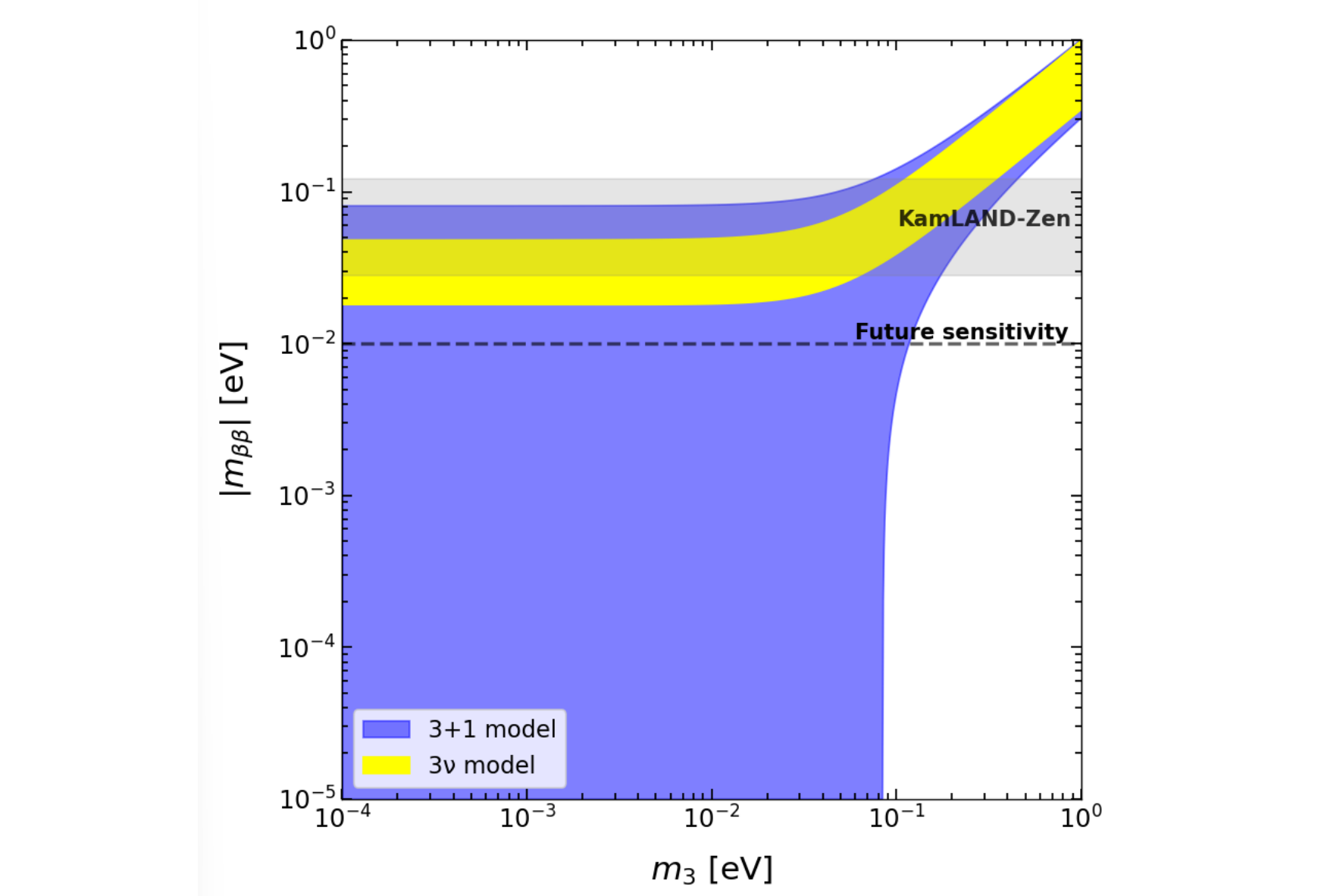}
	\caption{Effective Majorana mass $|m_{\beta\beta}|$ as a function of the lightest neutrino mass for NO (left) and IO (right). The blue and yellow bands correspond to the $3+1$ and $3\nu$ models, respectively. The 3+1 bands correspond to $\sin^2 2\theta_{14}=0.0665$ and $\Delta m_{41}^2=4.0~\mathrm{eV}^2$. The oscillation parameters are varied within the $3\sigma$ ranges in Table~\ref{tab:parameter_ranges}. The gray band shows the current KamLAND-Zen limit~\cite{KamLAND-Zen:2024eml}, and the dashed line indicates the future sensitivity $|m_{\beta\beta}|=0.01~\mathrm{eV}$~\cite{Ge:2025cky}.}
	\label{fig:sterilemlightest}
\end{figure}

Fig.~\ref{fig:sterilemlightest} compares the predicted $|m_{\beta\beta}|$ regions in the three-neutrino and 3+1 frameworks as functions of the lightest neutrino mass. The left and right panels
correspond to NO and IO, respectively. 
The bands are obtained by varying the oscillation parameters over the adopted $3\sigma$ intervals and each Majorana phase over $[0,2\pi)$.
The current KamLAND-Zen data imply $|m_{\beta\beta}|<0.028$--$0.122~\mathrm{eV}$, where the range reflects the uncertainty in the nuclear-matrix-element conversion~\cite{KamLAND-Zen:2024eml}.
Next-generation experiments such as nEXO~\cite{nEXO:2021ujk}, LEGEND-1000~\cite{Brugnera:2023zgw}, and CUPID~\cite{CUPID:2025faj} are expected to reach sensitivities near $m_{\beta\beta}\sim0.01$ eV. 
Under the light-neutrino-exchange hypothesis and for a specified NME, this sensitivity would probe the standard IO band and part of the NO parameter space.

In NO, the additional sterile contribution modifies the cancellation region and the boundaries of the effective-mass band.
This difference originates from the additional interference induced by the sterile-neutrino contribution, which modifies the funnel structure of the allowed region.
In IO, the 3+1 band extends to smaller values of $|m_{\beta\beta}|$ than the three-neutrino band.
This behavior results from the additional contribution $\mu_4e^{i\alpha_4}$, which can interfere destructively with the dominant active-neutrino contributions.
A null result covering the standard IO band could therefore be accommodated by sterile-induced cancellation, although it would not by itself establish the 3+1 scenario.

The qualitative changes shown in
Fig.~\ref{fig:sterilemlightest} are consistent with the earlier results shown in Ref.~\cite{Giunti:2015kza}. In particular, the additional term $\mu_4e^{i\alpha_4}$ modifies the cancellation region in both NO and IO and permits $|m_{\beta\beta}|$ to approach zero in IO, where the standard three-neutrino prediction has a nonzero lower bound.
The updated inputs, including the first JUNO results, refine the boundaries of these established cancellation regions.
The change is most apparent near the lower edge of the effective-mass band, where cancellation is sensitive to the relative magnitudes of $\mu_1$ and $\mu_2$ and hence to $\theta_{12}$ and $\delta m^2$.

\begin{figure}[t]
	\centering
	\includegraphics[width=0.49\linewidth]{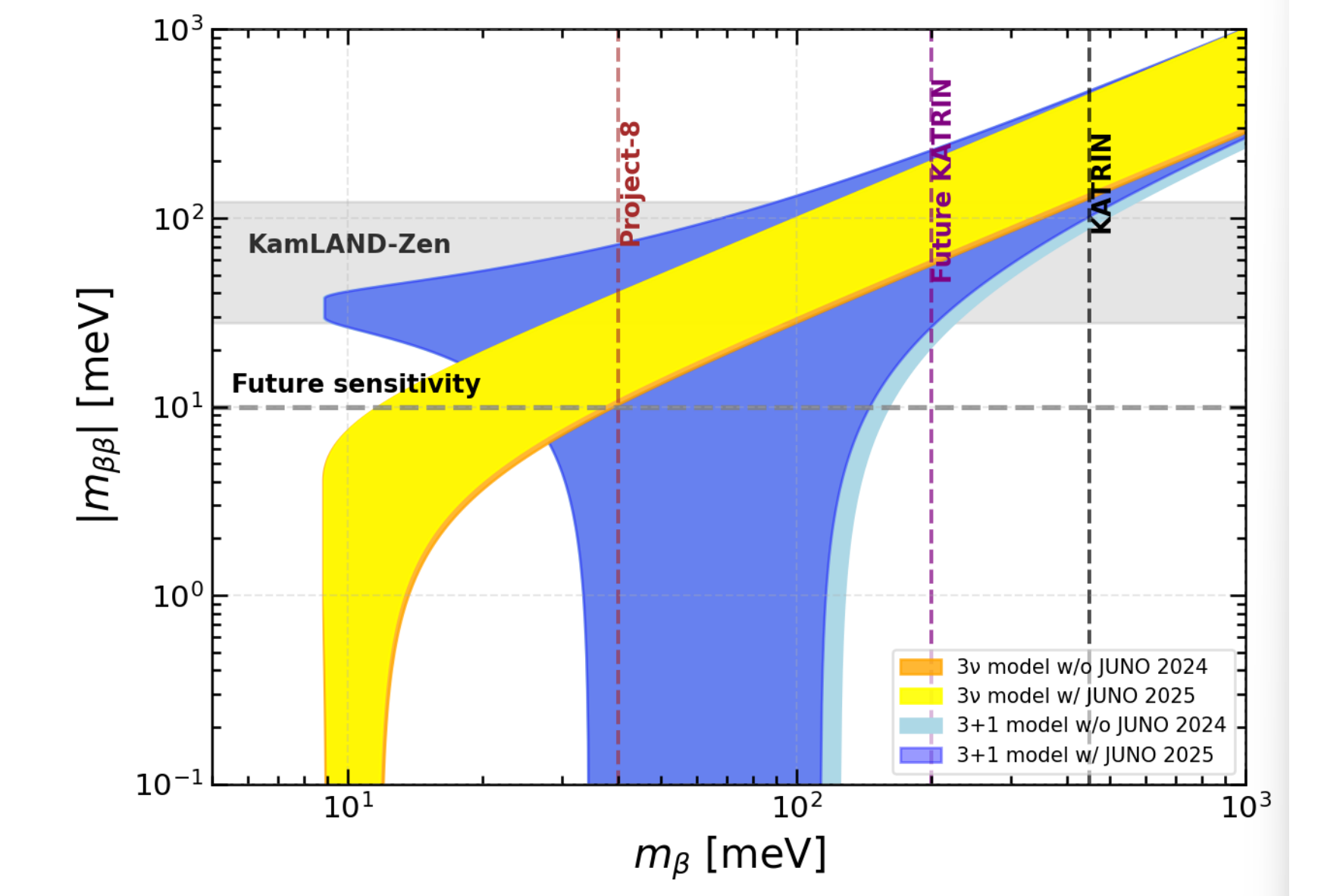}
	\hfill
	\includegraphics[width=0.49\linewidth]{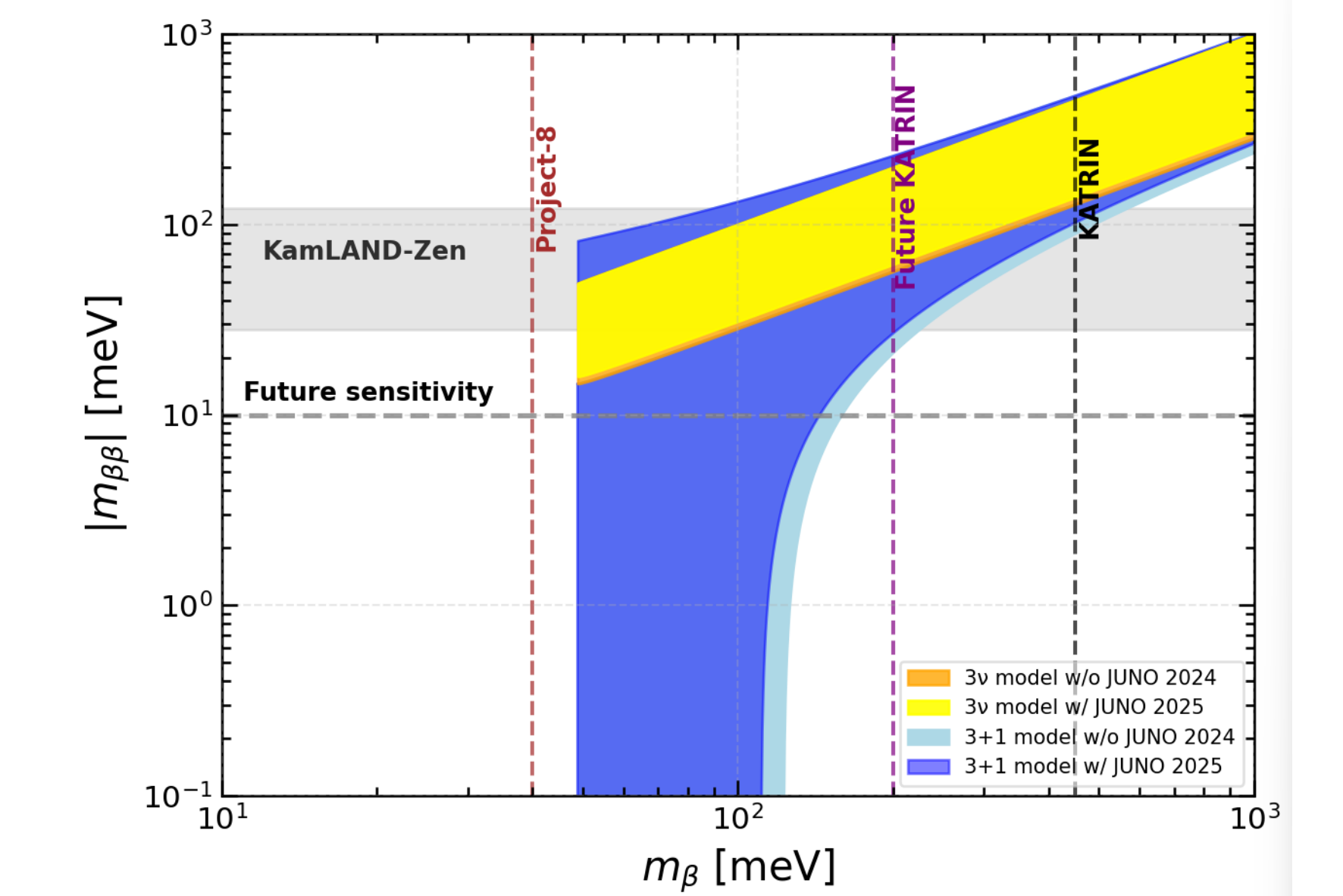}
	\caption{Allowed regions in the $m_\beta$--$|m_{\beta\beta}|$ plane in the $3\nu$ and 3+1 frameworks with the sterile-neutrino parameters as in Fig.~\ref{fig:sterilemlightest}. 
	The left and right panels correspond to NO and IO, respectively.
    Blue and yellow denote the $3+1$ and $3\nu$ predictions using the JUNO-informed 2025 oscillation inputs~\cite{Capozzi:2025ovi}, while light blue and orange show the corresponding pre-JUNO 2024 results~\cite{Capozzi:2025wyn}. The vertical dashed lines indicate the current KATRIN limit~\cite{KATRIN:2024cdt}, the KATRIN design sensitivity~\cite{KATRIN:2021uub}, and the Project~8 sensitivity~\cite{Project8:2022wqh}. }
	\label{fig:The31compare}
\end{figure}

We estimate the parametric uncertainty by varying the oscillation parameters within their $3\sigma$ ranges.
Fig.~\ref{fig:The31compare} shows the predicted $|m_{\beta\beta}|$--$m_\beta$ correlation obtained using the inputs in Table~\ref{tab:parameter_ranges}.
The figure also shows the current KATRIN limit ($m_{\beta} < 0.45\,\mathrm{eV}$ at 90\%\ C.L.)~\cite{KATRIN:2024cdt,Ge:2025cky}, the projected KATRIN sensitivity ($m_{\beta} < 0.2\,\mathrm{eV}$)~\cite{KATRIN:2021uub}, and the expected Project~8 sensitivity ($m_{\beta} < 0.04\,\mathrm{eV}$)~\cite{Project8:2022wqh}. 

The 3+1 and three-neutrino predictions differ most visibly in IO. This difference illustrates the complementary information carried by the two observables.
The bound $m_\beta\leq0.20\,\mathrm{eV}$ removes the high-mass part of both regions but leaves the sterile-induced cancellation at lower masses, particularly in IO. The two observables therefore probe distinct aspects of the neutrino mass spectrum and its interference structure.

Relative to the 2024 inputs, the JUNO-informed 2025 inputs reduce the oscillation-parameter contribution to the width of these regions.
The contraction is most visible near boundaries controlled by destructive interference, where changes in $\theta_{12}$ and $\delta m_{21}^2$ modify the relative sizes of the active-neutrino contributions. The JUNO-related update therefore narrows the predicted $|m_{\beta\beta}|$--$m_\beta$ regions, but does not remove the uncertainty associated with the unknown Majorana phases.

\subsection{Joint constraints on the sterile-neutrino parameter space}
\label{sec:Restrictsterileparameter}

We now vary $\theta_{14}$ and $\Delta m_{41}^{2}$ to determine the compatible sterile-neutrino parameter space.
Ref.~\cite{Jana:2024xmc} derived $0\nu\beta\beta$ constraints in the $(\sin^2 2\theta_{14},\Delta m_{41}^2)$ plane.
We follow a similar strategy, mapping limits on $m_{\beta\beta}$ and $m_{\beta}$ onto the sterile-neutrino parameter plane, while allowing a broader range of light-neutrino inputs. Rather than restricting the analysis to selected light-neutrino spectra and fixed relative-phase
configurations, we use prospective upper limits on
$|m_{\beta\beta}|$ and $m_{\beta}$ to determine compatible regions in the
$(\sin^2 2\theta_{14},\Delta m_{41}^2)$ plane. The lightest neutrino mass and the standard oscillation parameters are varied over their adopted
ranges, with the oscillation inputs updated to incorporate the first JUNO
results. 

We adopt $|m_{\beta\beta}|\leq0.01~\mathrm{eV}$ as a representative future $0\nu\beta\beta$ condition.
For the light-sector kinematic mass, we consider the two upper limits in Eq.~(\ref{eq:mbetalimit}).
The oscillation parameters are varied independently within the updated $3\sigma$ ranges listed in Table~\ref{tab:parameter_ranges}. We scan the lightest mass over $10^{-4}\,\mathrm{eV}\leq m_{\mathrm{lightest}}\leq1\,\mathrm{eV}$. Instead of explicitly sampling the Majorana phases, we minimize $|m_{\beta\beta}|$ analytically over all possible Majorana phases.
We obtain the minimum magnitude of the sum of the four complex contributions from Eq.~(\ref{eq:mbb_min}).
Complete cancellation is possible when the four vectors can form a closed polygon, which requires the largest magnitude not to exceed the sum of the other three. Otherwise, the minimum effective mass is the difference between the largest contribution and the sum of the remaining
ones.

For the same light-neutrino spectrum, the kinematic mass is calculated using Eq.~(\ref{eq:mbeta_light}). A sterile-neutrino parameter point is compatible if at least one sampled lightest mass and set of oscillation parameters satisfies both conditions
\begin{equation}
|m_{\beta\beta}|_{\min}
\leq
m_{\beta\beta}^{\mathrm{lim}}\,,
\qquad
m_\beta
\leq
m_{\beta,\,\mathrm{th}}\,.
\label{eq:III_B_acceptance}
\end{equation}

\begin{figure}[t]
	\centering
	\includegraphics[width=0.8\linewidth]{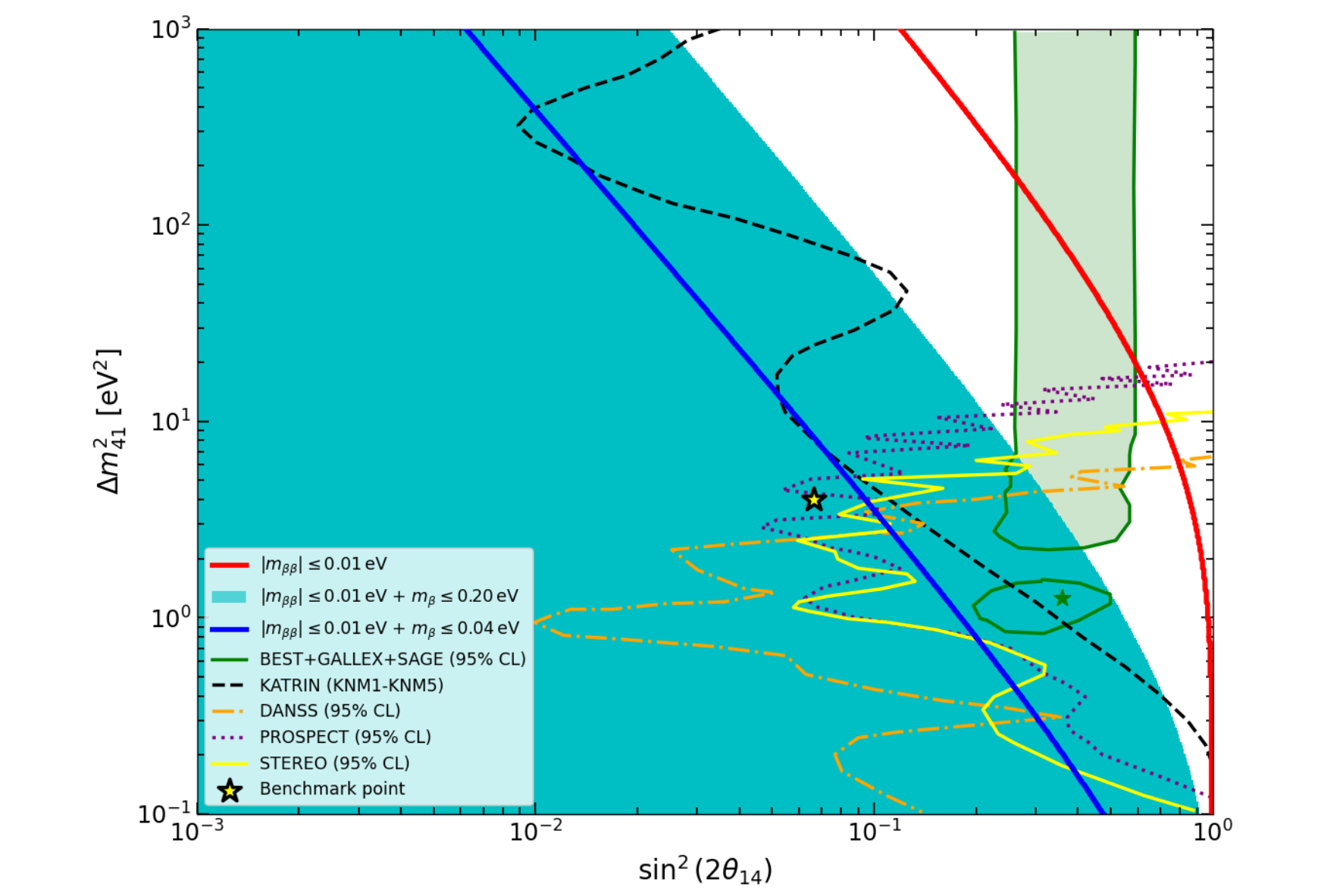}
	\caption{Allowed regions in the $(\sin^2 2\theta_{14},\Delta m_{41}^2)$ plane for NO, obtained with the JUNO-informed 2025 oscillation parameters. The region to the left of the red curve satisfies $|m_{\beta\beta}|\leq0.01~\mathrm{eV}$. The cyan region and the region to the left of the blue line additionally satisfy $m_\beta\leq0.20~\mathrm{eV}$ and $m_\beta\leq0.04~\mathrm{eV}$, respectively. The 95\% C.L. exclusion curves of KATRIN, DANSS, PROSPECT, and STEREO and the region favored by BEST+GALLEX+SAGE are taken from Ref.~\cite{KATRIN:2025lph}. The star denotes the benchmark point $\sin^2 2\theta_{14}=0.0665$, $\Delta m_{41}^2=4.0~\mathrm{eV}^2$.}
	\label{fig:sterilespaceNH}
\end{figure}

For NO, Fig.~\ref{fig:sterilespaceNH} shows the allowed sterile-neutrino parameter space after applying the $m_{\beta\beta}$ and $m_\beta$ conditions.
The region has a right-hand boundary that is approximately linear in the log--log plane at small mixing.
As $\theta_{14}$ increases, the boundary approaches a vertical asymptote.
The additional $m_\beta$ condition mainly restricts the lightest neutrino mass rather than directly limiting the sterile contribution. As a result, it reduces the active contribution available for cancellation and thereby shrinks the compatible sterile-neutrino region.
A tighter $m_\beta$ bound strengthens this restriction.
For NO, the pre-JUNO and JUNO-informed regions are nearly identical, showing that the improved precision of the solar oscillation parameters has only a minor effect on the projected sterile-neutrino compatibility region.


The shape of the compatibility region can be understood from the relative sizes of the four complex contributions.
Although $|U_{e1}|$ is the largest electron-row element, the mass $m_1$ is the smallest in NO. The dominant active contribution can therefore change across the scan.
For sufficiently small $\theta_{14}$, the sterile contribution remains subdominant, and the compatibility condition is controlled primarily by the active-neutrino terms.
Increasing $\theta_{14}$ enhances the sterile contribution.
Once $\mu_4$ exceeds each active contribution, it controls the right-hand boundary. Eq.~(\ref{eq:mbb_min}) then gives $\left | m_{\beta\beta} \right |_{\min}=\mu_{4} - (\mu_{1}+\mu_{2}+\mu_{3})$. 
The condition $|m_{\beta\beta}|_{\min}\leq m_{\beta\beta}^{\mathrm{lim}}$ therefore requires
\begin{equation}
s_{14}^{2}
\sqrt{m_1^{2}+\Delta m_{41}^{2}}
\leq
A_+
+
m_{\beta\beta}^{\mathrm{lim}}\,,
\label{eq:III_B_NO_condition}
\end{equation}
where
$A_+
\equiv
\mu_1+\mu_2+\mu_3$.
The right boundary in the sterile-dominated
region is consequently
\begin{equation}
\Delta m_{41}^{2}
\leq
\left(
\frac{A_++m_{\beta\beta}^{\mathrm{lim}}}{s_{14}^{2}}
\right)^{2}
-
m_1^{2}\,.
\label{eq:III_B_NO_boundary}
\end{equation}

In NO, the light-sector kinematic mass is
\begin{equation}
m_\beta^2
=
m_1^2
+
c_{13}^2s_{12}^2\delta m^2
+
s_{13}^2
\left(
\Delta m^2+\frac{\delta m^2}{2}
\right)\,.
\label{eq:III_B_mbeta_NO}
\end{equation}

The condition $m_\beta\leq m_{\beta\,,\mathrm{th}}$ then gives
\begin{equation}
0\leq m_1^2
\leq
m_{1,\max}^2
\equiv
m_{\beta\,,\mathrm{th}}^2-D_{\rm NO}\,,
\label{eq:III_B_m1max}
\end{equation}
where $D_{\rm NO}
\equiv
c_{13}^2s_{12}^2\delta m^2
+
s_{13}^2
\left(
\Delta m^2+\frac{\delta m^2}{2}
\right)$.
The largest active contribution allowed by this condition is obtained by evaluating
\begin{equation}
A_+^{\max}(m_{\beta\,,\mathrm{th}})
=
c_{14}^2c_{13}^2c_{12}^2m_{1,\max}
+
c_{14}^2c_{13}^2s_{12}^2m_2
+
c_{14}^2s_{13}^2m_3\,.
\label{eq:III_B_Aplus_explicit}
\end{equation}

In the sterile-dominated regime, the right-hand boundary of the NO compatibility region is approximately
\begin{equation}
\left(\Delta m_{41}^2\right)_{\rm R}^{\rm NO}
=
\left[
\frac{
A_+^{\max}(m_{\beta\,,\mathrm{th}})+ m_{\beta\beta}^{\mathrm{lim}}
}{
s_{14}^2
}
\right]^2
-
m_{1,\max}^2\,.
\label{eq:III_B_NO_right_boundary_exact}
\end{equation}

We have verified numerically that Eq.~(\ref{eq:III_B_NO_right_boundary_exact}) reproduces the  right-hand boundary in Fig.~\ref{fig:sterilespaceNH} and makes explicit how the light-sector mass bound restricts the sterile contribution.

\begin{figure}[t]
	\centering
	\includegraphics[width=0.8\linewidth]{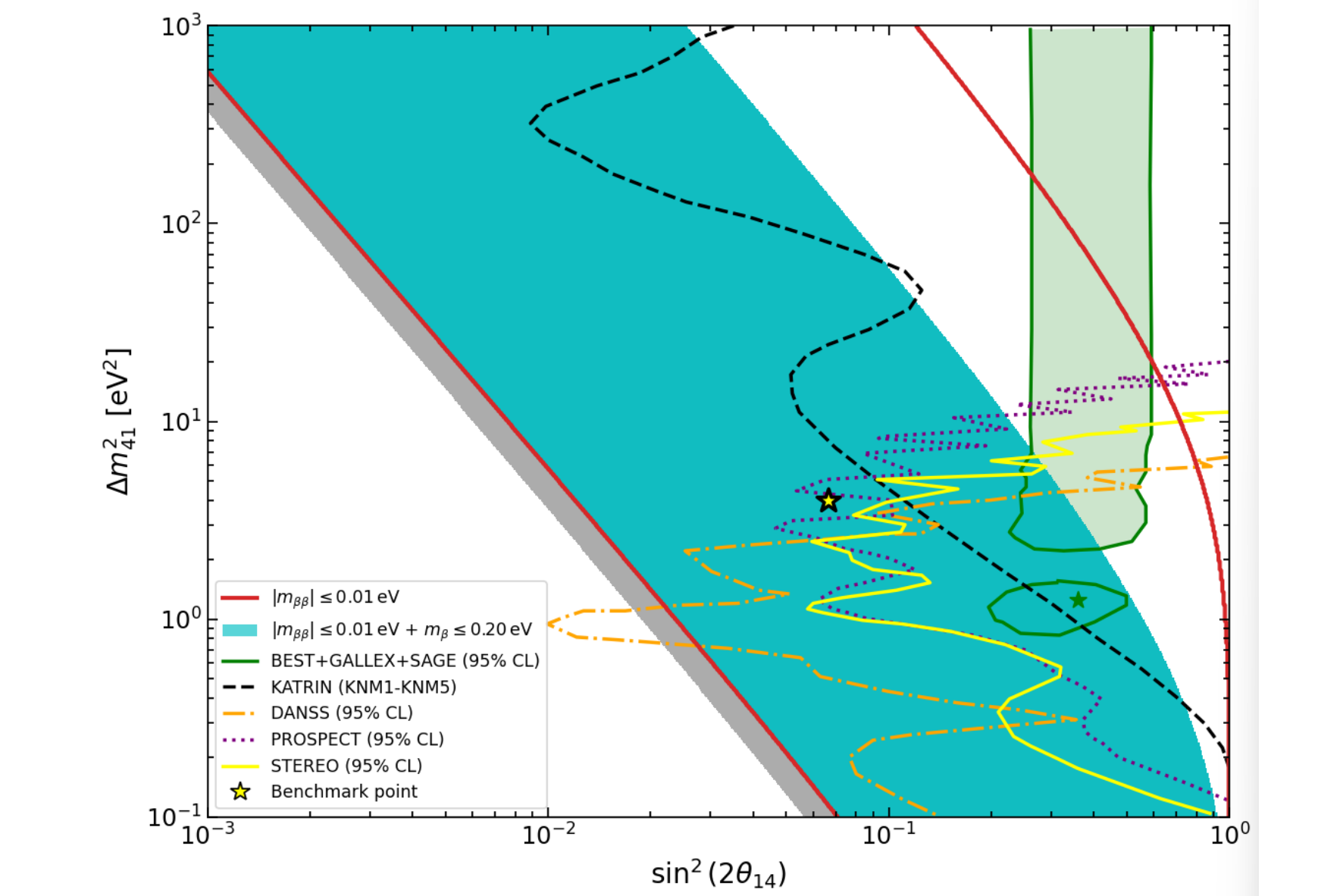}
	\caption{Same as Fig.~\ref{fig:sterilespaceNH} for IO. The region between the two red curves satisfies $|m_{\beta\beta}|\leq0.01~\mathrm{eV}$, and the cyan region additionally satisfies $m_\beta\leq0.20~\mathrm{eV}$. The gray region shows the corresponding result obtained with the pre-JUNO 2024 oscillation inputs. There is no allowed region for $m_\beta\leq0.04~\mathrm{eV}$.}
	\label{fig:sterilespaceIH}
\end{figure}

For IO, no scanned point satisfies both $|m_{\beta\beta}|\leq0.01\,\mathrm{eV}$ and $m_\beta\leq0.04\,\mathrm{eV}$ under the adopted light-sector convention and sterile-neutrino constraints.
We therefore restrict the following discussion to the case of $|m_{\beta\beta}|\le 0.01\,\mathrm{eV}$ and $m_{\beta}\le 0.20\,\mathrm{eV}$. Fig.~\ref{fig:sterilespaceIH} shows the allowed sterile-neutrino parameter space for IO.
The IO band has both lower and upper boundaries. At small sterile mixing, $\mu_1$ is the largest active contribution.
For sufficiently small $\theta_{14}$, $\mu_1$ remains dominant, and Eq.~(\ref{eq:mbb_min}) gives $\left |m_{\beta\beta}\right |_{\min}=\mu_{1}-(\mu_{2}+\mu_{3}+\mu_{4})$.
The condition $|m_{\beta\beta}|_{\min}\leq m_{\beta\beta}^{\mathrm{lim}}$ therefore requires
\begin{equation}
s_{14}^2
\sqrt{m_1^2+\Delta m_{41}^2}
\geq
A_- - m_{\beta\beta}^{\mathrm{lim}}\,,
\end{equation}
where
$A_-
\equiv
\mu_1-\mu_2-\mu_3$.
The corresponding lower boundary is
\begin{equation}
\left(\Delta m_{41}^2\right)_{\mathrm{L}}^{\mathrm{IO}}
=
\left[
\frac{
\max(A_--m_{\beta\beta}^{\mathrm{lim}},0)
}{
s_{14}^2
}
\right]^2
-
m_1^2\,.
\label{eq:III_B_IO_left_boundary}
\end{equation}

As $\theta_{14}$ or $m_4$ increases, the sterile contribution eventually becomes the largest term. In this regime,
\begin{equation}
|m_{\beta\beta}|_{\min}
=
\mu_4-(\mu_1+\mu_2+\mu_3)
=
\mu_4-A_+\,.
\label{eq:III_B_IO_mu4_dominant}
\end{equation}

The condition
$|m_{\beta\beta}|_{\min}\leq m_{\beta\beta}^{\mathrm{lim}}$ then gives
\begin{equation}
s_{14}^2
\sqrt{m_1^2+\Delta m_{41}^2}
\leq
A_+
+
m_{\beta\beta}^{\mathrm{lim}}\,.
\end{equation}

The corresponding upper boundary is 
\begin{equation}
\left(\Delta m_{41}^2\right)_{\mathrm{R}}^{\mathrm{IO}}
=
\left[
\frac{
A_++m_{\beta\beta}^{\mathrm{lim}}
}{
s_{14}^2
}
\right]^2
-
m_1^2\,.
\label{eq:III_B_IO_right_boundary}
\end{equation}

Eqs.~(\ref{eq:III_B_IO_left_boundary}) and (\ref{eq:III_B_IO_right_boundary}) explain the two sides of the IO band: the sterile amplitude must be large enough to cancel the active residual but small enough to be canceled by the full active amplitude.
Compared with the 2024 inputs, the JUNO-informed inputs reduce the uncertainty in the left boundary while leaving the right boundary nearly unchanged.
The improved precision of $\theta_{12}$ and $\delta m^2$ reduces the uncertainty in the active residual $A_-$.
The left-hand boundary is set by cancellation of this residual and is therefore sensitive to the solar oscillation parameters.
The right-hand boundary instead depends on the maximum active sum $A_+$ available to cancel the dominant sterile term and is less sensitive to these oscillation inputs.
The agreement between the numerical boundaries and the analytic conditions provides a consistency check of the scan.
The JUNO update therefore primarily narrows the uncertainty on the active-residual side of the IO band.

\subsection{Sterile Majorana phase and mass-dependent cancellation structure}
\label{sec:alpha4}

The preceding analysis identifies the sterile masses and mixing angles compatible with the assumed effective-mass bounds after varying the Majorana phases. Because compatibility depends on interference among the four mass-eigenstate contributions, the surviving points also restrict the relative phases.
We now determine the compatible values of $\alpha_4$ while allowing $\alpha_2$ and $\alpha_3$ to vary over their full ranges. The exact-cancellation condition $m_{\beta\beta}=0$ in the 3+1 framework has been recently studied in Ref.~\cite{Chambyal:2026fnq}.
Here we impose the finite bound $|m_{\beta\beta}|\leq0.01~\mathrm{eV}$ and determine the corresponding compatible range of $\alpha_4$. 
In particular, we use a complex-plane geometric analysis to derive the compatibility condition for $\alpha_4$ and explain how the allowed phase regions depend on the sterile mass, the mass ordering, and the assumed upper limit on $|m_{\beta\beta}|$.

\begin{figure}[t]
	\centering
	\includegraphics[width=0.49\linewidth]{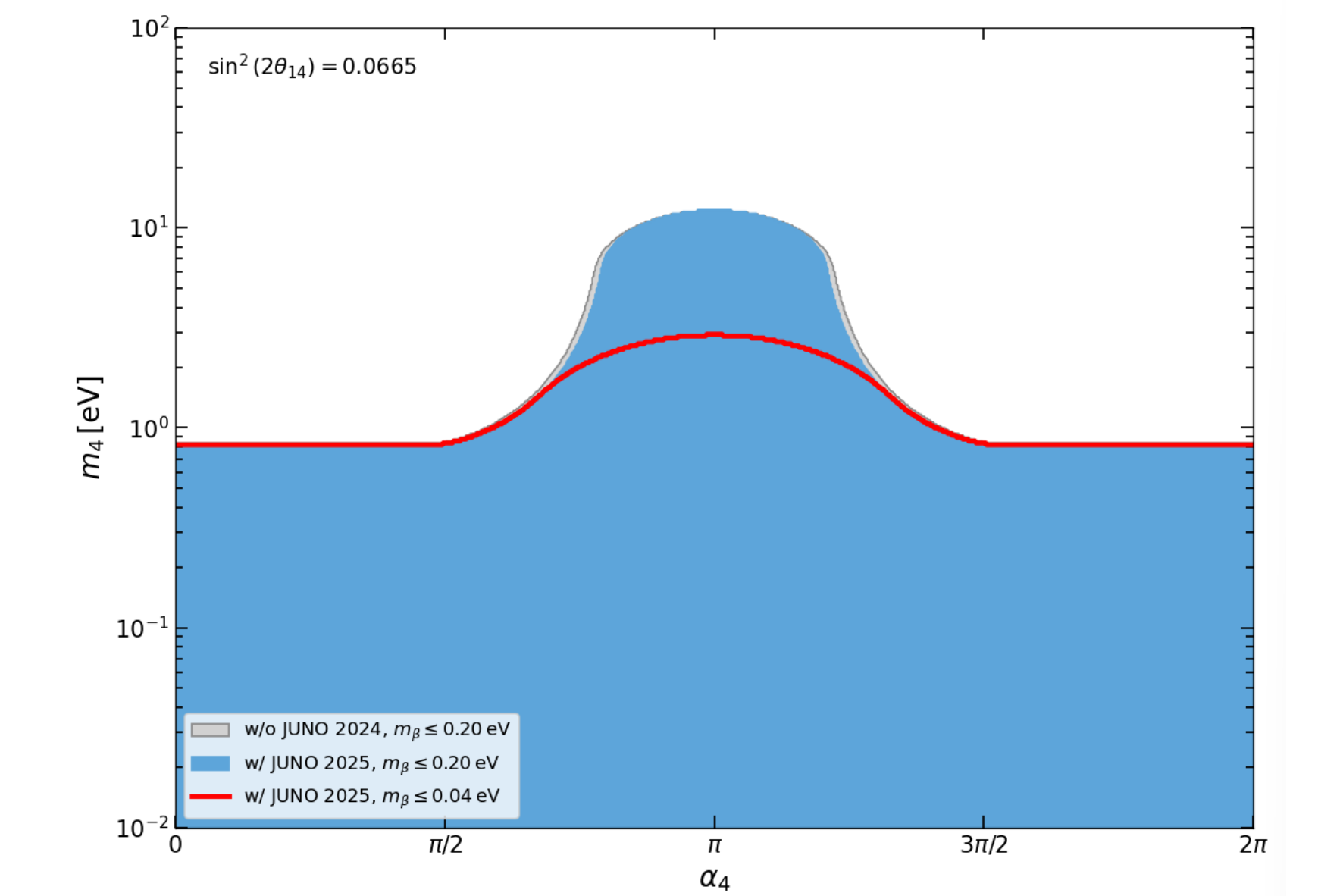}
	\hfill
	\includegraphics[width=0.49\linewidth]{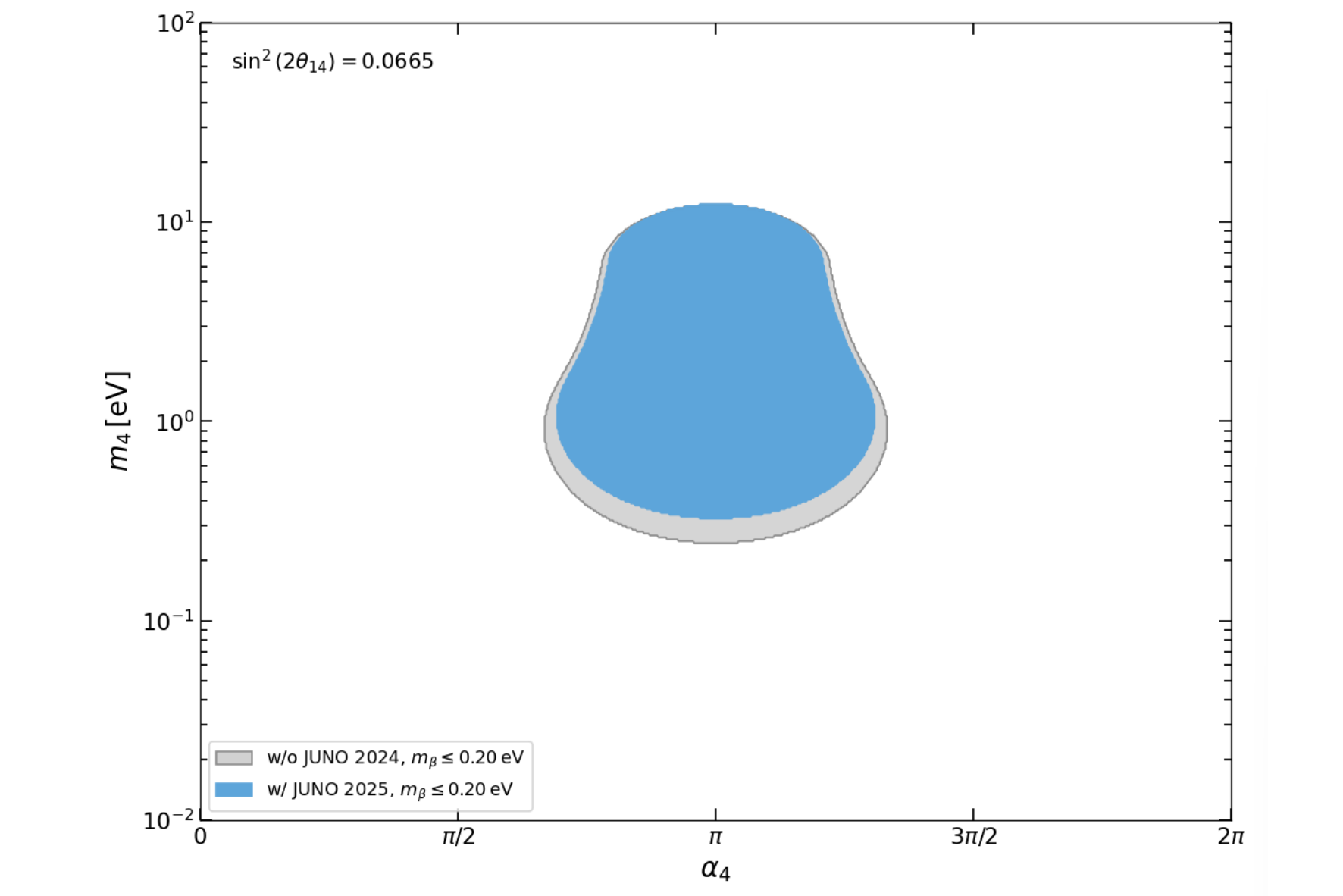}
	\caption{Allowed regions in the $(\alpha_4,m_4)$ plane for NO (left) and IO (right) with $\sin^2 2\theta_{14}=0.0665$ and $|m_{\beta\beta}|\leq0.01~\mathrm{eV}$. The blue and gray regions are obtained with $m_\beta\leq0.20~\mathrm{eV}$ using the 2025 and 2024 oscillation parameters, respectively. In the left panel, the region below the red curve satisfies $m_\beta\leq0.04~\mathrm{eV}$.
    }
	\label{fig:alpha4&m4}
\end{figure}

We first examine the correlation between $\alpha_4$ and the sterile-neutrino mass. Fig.~\ref{fig:alpha4&m4} shows the compatible regions in the $(\alpha_4,m_4)$ plane
for fixed $
    \sin^2(2\theta_{14})=0.0665\,,
$
with the standard oscillation parameters varied within their adopted $3\sigma$ ranges. The condition
$|m_{\beta\beta}|\leq0.01~{\rm eV}$ is imposed together with the adopted light-sector mass condition.
The resulting phase structure depends strongly on both $m_4$ and the mass
ordering. In NO, a broad range of $\alpha_4$ is compatible at small $m_4$, whereas larger masses favor phases near $\pi$. Above a limiting mass, the active amplitude can no longer compensate the sterile contribution and no compatible phase remains. 
In IO, the allowed region of $\alpha_4$ is more restricted,
with $\alpha_4$ confined to a narrower interval around $\pi$.
The JUNO-informed inputs visibly modify the IO region but have little effect on the NO region.

\begin{figure}[t]
	\centering
\subfigure[]{\includegraphics[width=0.32\linewidth]{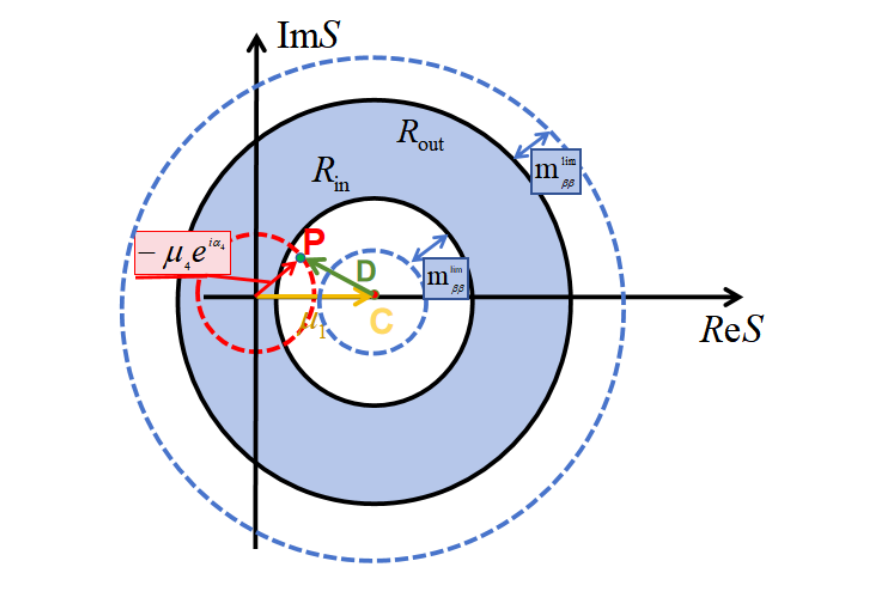}}
	\hfill
    \subfigure[]{\includegraphics[width=0.32\linewidth]{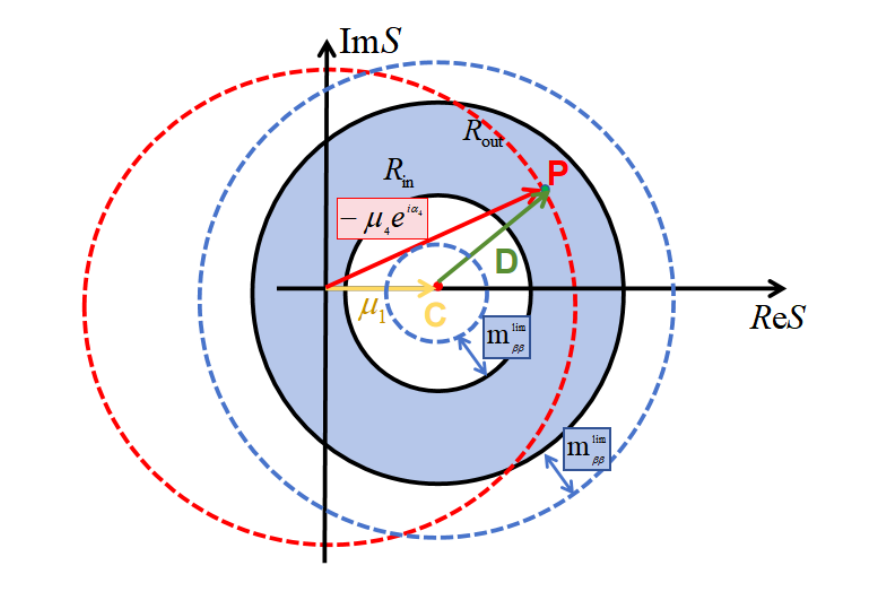}}
     \subfigure[]{\includegraphics[width=0.32\linewidth]{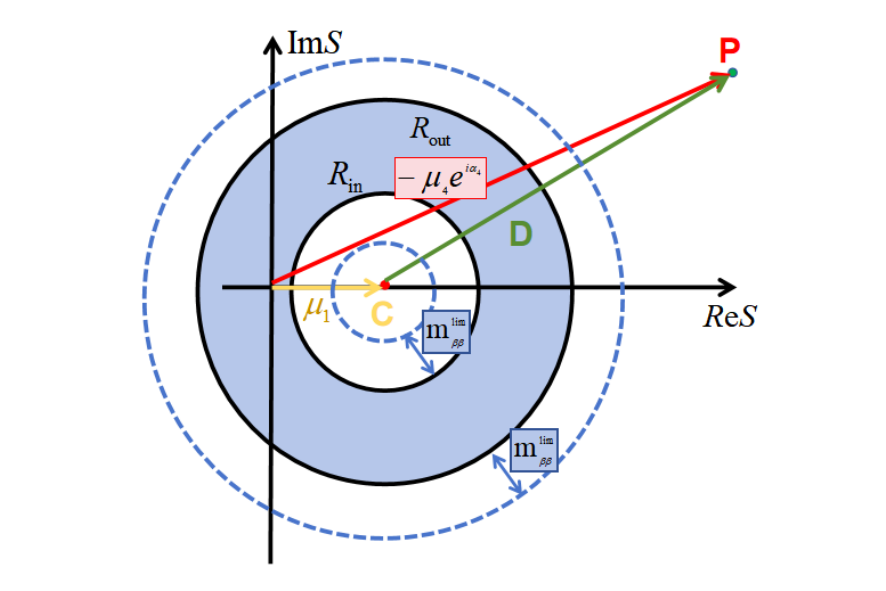}}
	\hfill
    \subfigure[]{\includegraphics[width=0.32\linewidth]{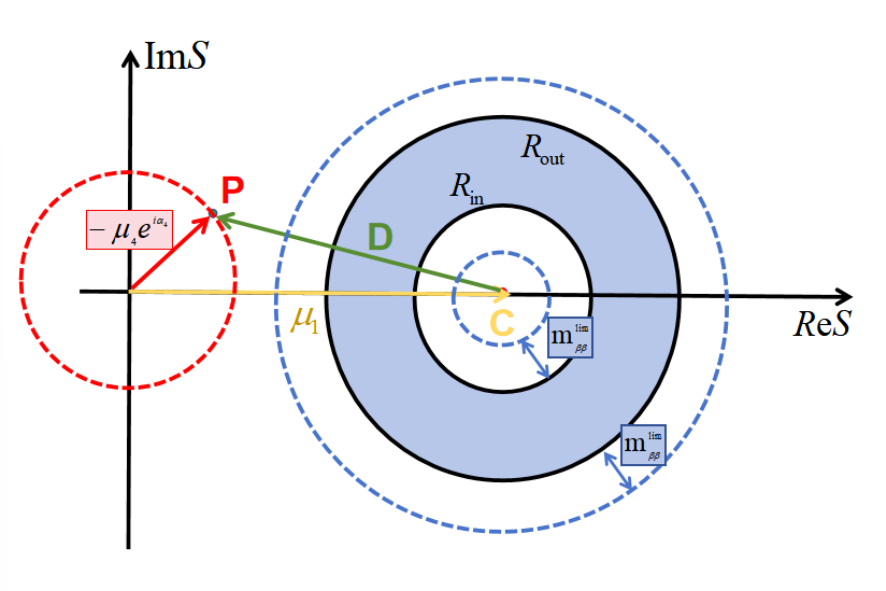}}
	\hfill
    \subfigure[]{\includegraphics[width=0.32\linewidth]{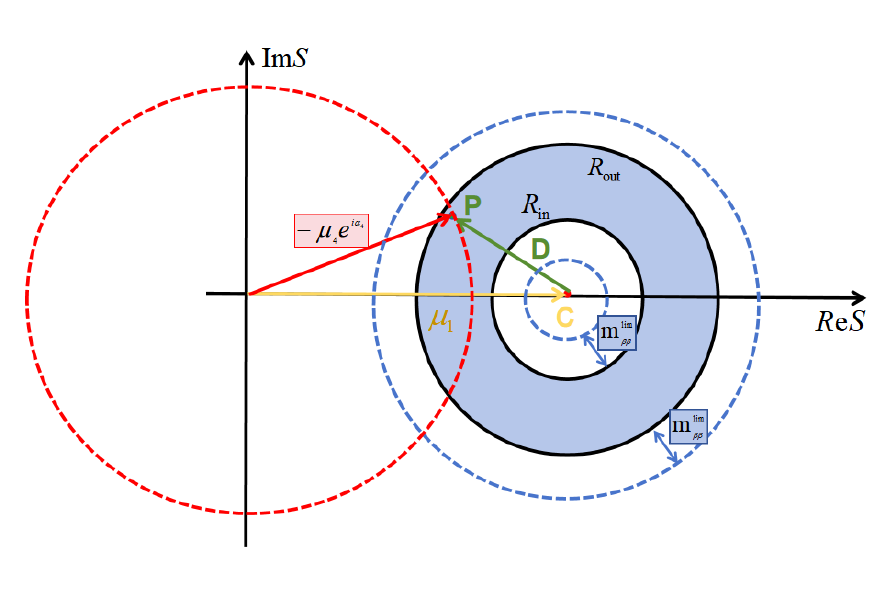}}
     \subfigure[]{\includegraphics[width=0.32\linewidth]{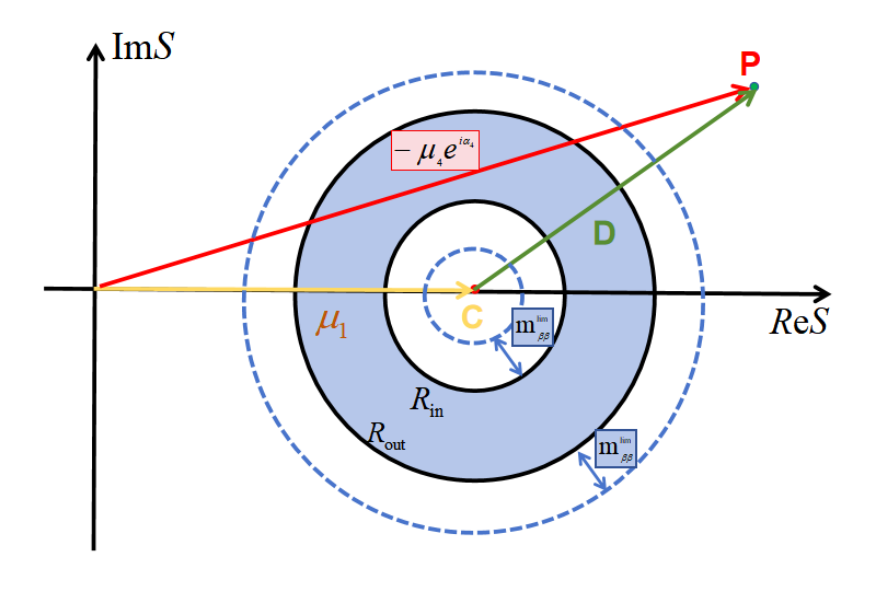}}
	\caption{Geometric representation of $m_{\beta\beta}$ in the complex plane for small, intermediate, and large $m_4$ (from left to right) in NO [(a)--(c)] and IO [(d)--(f)]. The shaded annulus, centered at $C=(\mu_1,0)$ with radii $R_{\rm in}$ and $R_{\rm out}$, contains the active contribution $S$ for all values of $\alpha_2$ and $\alpha_3$. The red dashed circle is traced by $P=-\mu_4e^{i\alpha_4}$ as $\alpha_4$ varies, and the blue dashed circles of radii $R_{\rm in}-m_{\beta\beta}^{\rm lim}$ and $R_{\rm out}+m_{\beta\beta}^{\rm lim}$ delimit the allowed positions of $P$ given by Eq.~(\ref{eq:DRange}). The figures are not to scale.}
	\label{fig:geometric}
\end{figure}

The origin of the different structures in Fig.~\ref{fig:alpha4&m4} can be understood geometrically from the complex-vector representation of $m_{\beta\beta}$, as shown in Fig.~\ref{fig:geometric}.
We separate the active-neutrino contribution in Eq.~(\ref{eq:sterilembetabeta}) as $S=\mu_1+\mu_2e^{i\alpha_2}+\mu_3e^{i\alpha_3}$, where
$\mu_k=m_k|U_{ek}|^2$. 
At fixed masses and mixing angles, varying $\alpha_2$ and $\alpha_3$ makes $S$ sweep out an annulus centered at $(\mu_1,0)$. Its inner and outer radii are
\begin{equation}
	\label{eq:R}
	R_{\mathrm{in}}=\left |\mu_{2}-\mu_{3}  \right | \,,
	R_{\mathrm{out}}=\mu_{2}+\mu_{3} \,. 
\end{equation}

The effective Majorana mass can then be written as
\begin{equation}
	\label{eq:sterile}
	\left | m_{\beta\beta} \right |=\left | S + \mu_{4}e^{i\alpha_{4}} \right | \,.
\end{equation}

For an assumed bound $|m_{\beta\beta}|\leq m_{\beta\beta}^{\mathrm{lim}}$, a given $\alpha_4$ is compatible if and only if the disk of radius $m_{\beta\beta}^{\mathrm{lim}}$ centered at $P=-\mu_4e^{i\alpha_4}$ intersects the active annulus.
The distance between $P$ and the center of the active annulus is
\begin{equation}
	\label{eq:D}
	D=\left | \mu_{1}+\mu_{4}e^{i\alpha_{4}} \right | =\sqrt{\mu_{1}^2+\mu_{4}^2+2\mu_{1}\mu_{4}\cos\alpha_{4}} \,.
\end{equation}

The minimum distance from $P$ to the set of allowed active amplitudes, denoted by $\operatorname{dist}(P,S)$, is
\begin{equation}
	\label{eq:dist}
	\operatorname{dist}(P\,,S)=\left\{\begin{matrix}
		R_{\mathrm{in}}-D\,,\,\,\,D<R_{\mathrm{in}}\,, \\
		0\,,\,\,\,R_{\mathrm{in}}\le D\le R_{\mathrm{out}}\,,\\
		D-R_{\mathrm{out}}\,,\,\,\,D>R_{\mathrm{out}}\,.
	\end{matrix}\right.
\end{equation}

Requiring this distance to be at most $m_{\beta\beta}^{\mathrm{lim}}$ gives
\begin{equation}
	\max(0,\,R_{\mathrm{in}}-m_{\beta\beta}^{\mathrm{lim}}) \le D\le R_{\mathrm{out}}+m_{\beta\beta}^{\mathrm{lim}} \,.
\label{eq:DRange}
\end{equation}

At fixed masses and mixing angles, this condition accounts exactly for the full ranges of $\alpha_2$ and $\alpha_3$, leaving a direct constraint on $\alpha_4$ through $D$. The compatible phases correspond to the portions of the circle traced by $P$ that lie within a distance $m_{\beta\beta}^{\mathrm{lim}}$ of the active annulus.

Fig.~\ref{fig:geometric} illustrates this condition for representative small,
intermediate, and large values of $m_4$ in NO and IO. The geometry explains the mass dependence of the compatible regions in Fig.~\ref{fig:alpha4&m4}.
In the small-$m_4$ NO examples, the sterile amplitude is small enough to be accommodated by the active contribution over a broad phase range.
The point $P$ traces a small circle, much of which lies within a distance $m_{\beta\beta}^{\mathrm{lim}}$ of the active annulus.
The resulting restriction on $\alpha_4$ is weak.
As $m_4$ increases, the sterile amplitude becomes comparable to the active amplitude.
The circle traced by $P$ grows, and only part of it satisfies Eq.~(\ref{eq:DRange}). 
In the configurations shown, the compatible phases consequently concentrate around $\alpha_4=\pi$. 
At this phase,
\begin{equation}
    D_{\min}=|\mu_1-\mu_4|\,,
\end{equation}
which gives the minimum distance between $P$ and the center of the active annulus.
If $P$ lies inside its central hole, the lower inequality in Eq.~(\ref{eq:DRange}) also matters; phases displaced from $\pi$ can then be required. Near the large-$m_4$ boundary, it is the outer inequality that selects phases close to $\pi$.
At still larger $m_4$, the sterile amplitude dominates. Near the upper compatible mass, only phases close to $\pi$ keep $P$ within the required distance of the active annulus. Once
\begin{equation}
    |\mu_4-\mu_1|>R_{\rm out}+m_{\beta\beta}^{\mathrm{lim}} \,,
\end{equation}
even the closest point on the circle lies beyond the outer allowed radius, so no phase satisfies the assumed bound.

The IO region follows from the same geometry.
In IO, $\nu_1$ and $\nu_2$ give sizable contributions even as the lightest mass tends to zero, while the $\nu_3$ contribution remains small. In the relevant IO region, the active annulus does not reach the origin.
A sufficiently small sterile amplitude cannot cancel the nonzero active residual to the required precision. Compatibility requires a sterile amplitude large enough to compensate this residual and a phase that gives destructive interference.
The compatible IO phases therefore lie near $\pi$ and become restricted again as $m_4$ approaches the upper boundary.

At $\alpha_4=\pi$, the sterile amplitude $\mu_4e^{i\alpha_4}$ points along the negative real axis.
Its negative, $P=-\mu_4e^{i\alpha_4}$, lies on the positive real axis, as shown in the geometric construction.
The circle in Fig.~\ref{fig:geometric} thus tracks the negative of the sterile amplitude, not the amplitude itself. Increasing $m_4$ changes both the sterile amplitude and the range of phases that permits sufficient cancellation. 

\begin{figure}[t]
	\centering
	\includegraphics[width=0.49\linewidth]{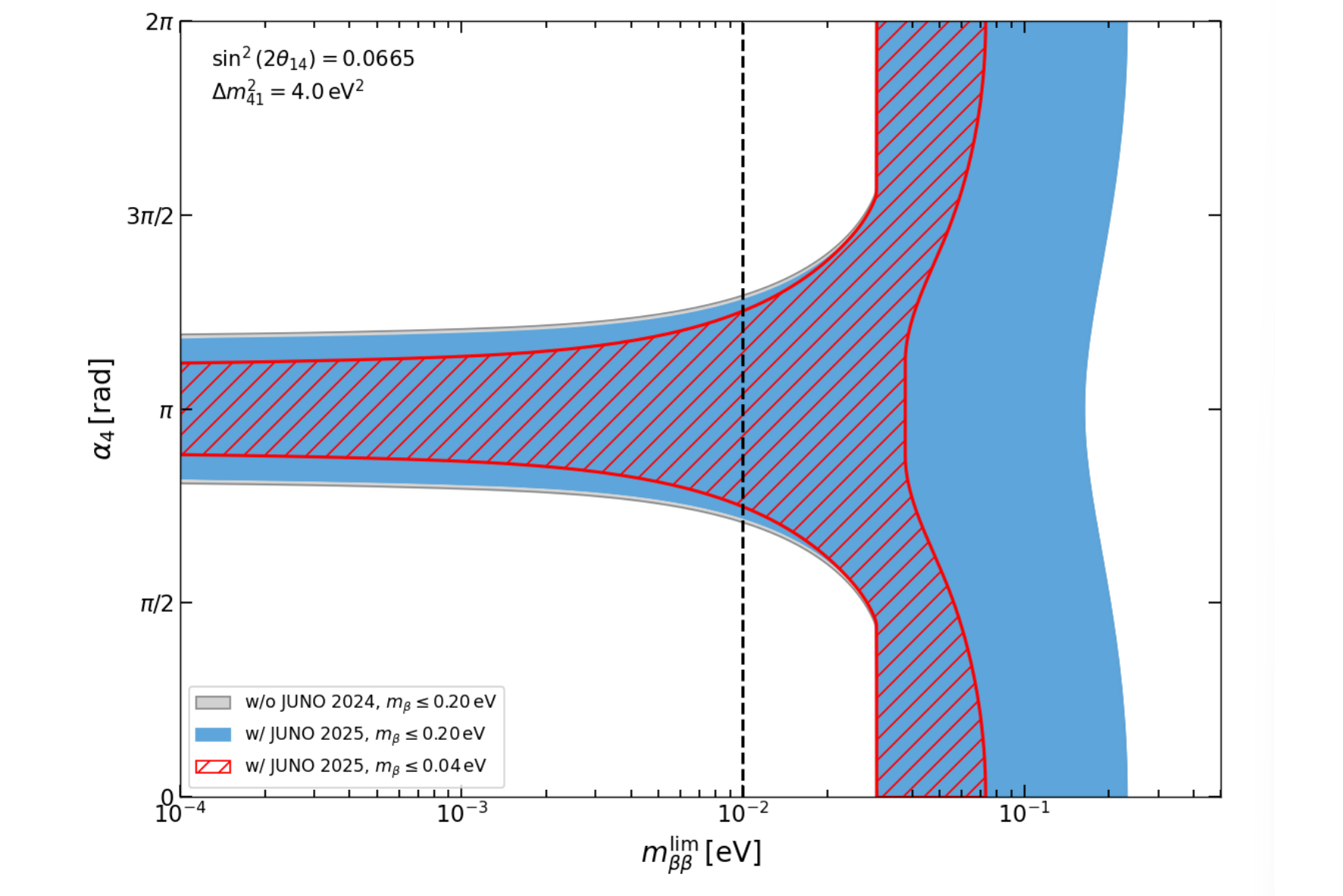}
	\hfill
	\includegraphics[width=0.49\linewidth]{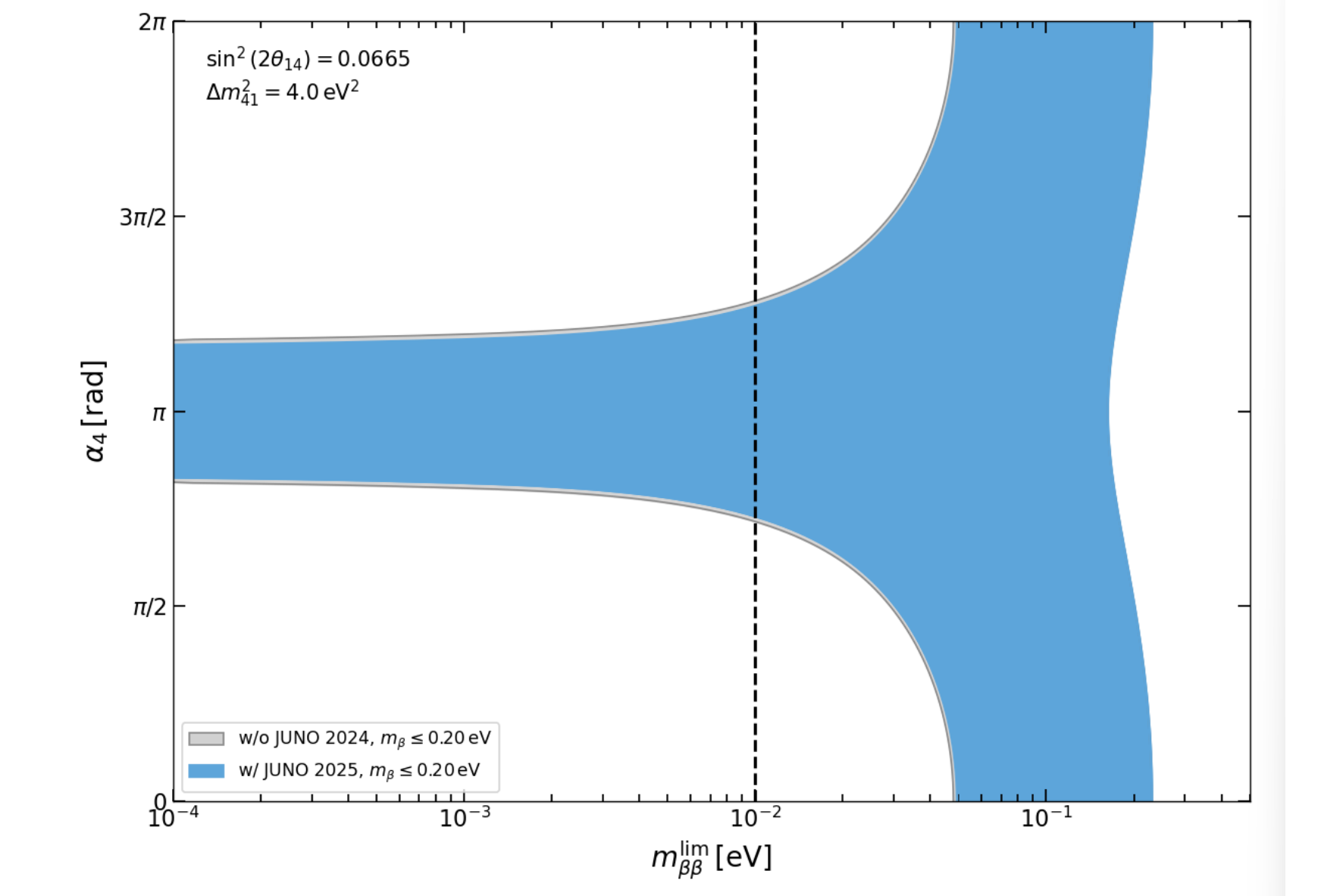}
	\caption{Allowed regions in the $(m_{\beta\beta}^{\rm lim},\alpha_4)$ plane for NO (left) and IO (right) with $\sin^2 2\theta_{14}=0.0665$ and $\Delta m_{41}^2=4.0~\mathrm{eV}^2$. The blue and gray regions are obtained with $m_\beta\leq0.20~\mathrm{eV}$ using the 2025 and 2024 oscillation parameters, respectively. In the left panel, the red hatched region is obtained with $m_\beta\leq0.04~\mathrm{eV}$. The vertical dashed line indicates $m_{\beta\beta}^{\rm lim}=0.01~\mathrm{eV}$.}
	\label{fig:alphabeta}
\end{figure}

We next hold the sterile parameters fixed and vary the assumed upper limit on $|m_{\beta\beta}|$.
Fig.~\ref{fig:alphabeta} shows the compatible $\alpha_4$ intervals as a function of the assumed upper limit on $|m_{\beta\beta}|$ at the benchmark sterile-neutrino point
\begin{equation}
    \sin^2(2\theta_{14})=0.0665,
    \qquad
    \Delta m_{41}^{2}=4.0~{\rm eV}^{2}\,.
\end{equation}

A weaker upper limit $m_{\beta\beta}^{\mathrm{lim}}$ permits a broad phase interval, whereas a tighter limit restricts the compatible phases. 
In particular, configurations satisfying the prospective sensitivity near $m_{\beta\beta}^{\mathrm{lim}}=0.01~{\rm eV}$ are concentrated around the $\alpha_4\simeq\pi$ region.

This behavior follows directly from Eq.~(\ref{eq:sterile}).
For a fixed active amplitude $S=|S|e^{i\phi_S}$, we write $\alpha_4=\phi_S+\pi+\sigma$, where $\sigma$ measures the departure from exact antialignment.
Substitution into Eq.~(\ref{eq:sterile}) gives
\begin{equation}
	\label{eq:Mbetabeta}
	\left | m_{\beta\beta} \right |=\left | \left | S \right |-\mu_{4}e^{i\sigma}   \right | =\sqrt{(\left | S \right |-\mu_{4})^2+4\left | S \right |\mu_{4}\sin^2\frac{\sigma}{2}  } \,.
\end{equation}

Varying $\sigma$ at fixed $|S|$ and $\mu_4$ gives the extrema
\begin{equation}
    M_{\min}
    =
    \big||S|-\mu_4\big|,
    \qquad
    M_{\max}
    =
    |S|+\mu_4 \,.
\end{equation}

For $M_{\min}\leq m_{\beta\beta}^{\mathrm{lim}}\leq M_{\max}$, the bound $|m_{\beta\beta}|\leq m_{\beta\beta}^{\mathrm{lim}}$ requires
\begin{equation}
    \sin^2\left(\frac{\sigma}{2}\right)
    \leq
    \frac{
    (m_{\beta\beta}^\mathrm{lim})^2-(|S|-\mu_4)^2
    }{
    4|S|\mu_4
    } \,.
    \label{eq:sigmabound}
\end{equation}

The corresponding range of $\alpha_4$ is
\begin{equation}
	\label{eq:alpha4}
	\phi _{S}+\pi-2\arcsin \sqrt{\frac{(m_{\beta\beta}^\mathrm{lim})^2-(\left |S  \right |-\mu_{4})^2 }{4\left | S \right |\mu_{4} }
	} \le \alpha_{4} \le \phi _{S}+\pi+2\arcsin \sqrt{\frac{(m_{\beta\beta}^\mathrm{lim})^2-(\left |S  \right |-\mu_{4})^2 }{4\left | S \right |\mu_{4} }
	}\,.
\end{equation}

The full regions in Fig.~\ref{fig:alphabeta} can be obtained by allowing the active amplitude to vary with the light-sector masses, mixing parameters, and phases. As the upper limit on $m_{\beta\beta}$ decreases, the compatible range of $\alpha_4$ narrows, while cancellation regions near $\alpha_4\simeq\pi$ remain viable for the benchmark shown. 
The pre-JUNO and JUNO-informed regions largely overlap for both orderings, showing that the oscillation update has only a limited effect on these phase--sensitivity correlations.

\section{Conclusion}
\label{sec:conclusion}

We have studied the combined constraints from $0\nu\beta\beta$ and beta decay on eV-scale sterile Majorana neutrinos in the 3+1 framework. Using updated oscillation parameters, including the first JUNO results, we have obtained the allowed regions in the $(\sin^2 2\theta_{14},\Delta m_{41}^2)$ plane for both mass orderings. We have also examined how the allowed values of the sterile Majorana phase $\alpha_4$ depend on the sterile mass and the sensitivity of $0\nu\beta\beta$ experiments.

We find that the beta-decay mass bound strengthens the constraints from $0\nu\beta\beta$ alone. The reason is that an upper limit on $m_\beta$ restricts the light-neutrino masses and reduces the maximum active contribution available to cancel the sterile term in $m_{\beta\beta}$. We derive analytic expressions for the boundaries of the allowed regions and illustrate the improvement for $|m_{\beta\beta}|\leq0.01\,\mathrm{eV}$ with $m_\beta\leq0.20\,\mathrm{eV}$ and $0.04\,\mathrm{eV}$. In inverted ordering, the sterile contribution must be large enough to cancel the nonzero active contribution, but it cannot exceed the amount allowed by the beta-decay bound. This leads to both a lower and an upper boundary. 
Under the adopted light-sector mass definition and oscillation parameter ranges, no IO solution satisfies the illustrative bound $m_\beta\leq0.04\,\mathrm{eV}$, since the minimum effective beta-decay mass in IO is approximately $0.048\,\mathrm{eV}$.

We use a geometric representation of $m_{\beta\beta}$ in the complex plane to explain the constraints on $\alpha_4$. Allowing the other two Majorana phases to vary, we derive an analytic condition for the allowed $\alpha_4$ range at a finite upper limit on $|m_{\beta\beta}|$. For the mixing considered, this explains why normal ordering permits a broad phase range at small sterile masses, while inverted ordering favors $\alpha_4$ near $\pi$. As the sterile mass approaches its upper allowed value, stronger cancellation is required and the phase range narrows. A more stringent $|m_{\beta\beta}|$ limit further restricts this range. The geometric analysis thus gives a physical explanation of the phase dependence seen in the numerical results.

The first JUNO results mainly reduce the uncertainty near the cancellation boundary in inverted ordering; their effect on the overall sterile-neutrino parameter space is modest. The main improvement comes from combining the two effective-mass constraints. Our results show how future beta-decay and $0\nu\beta\beta$ measurements can jointly constrain the sterile-neutrino mass, mixing, and Majorana phase.

\acknowledgments
We thank Yu-Feng Li for helpful discussions. J.~L. is supported by the National Natural Science Foundation of China under Grant No.~12275368.

\bibliography{references}

@article{Bilenky:2014uka,
    author = "Bilenky, S. M. and Giunti, C.",
    title = "{Neutrinoless Double-Beta Decay: a Probe of Physics Beyond the Standard Model}",
    eprint = "1411.4791",
    archivePrefix = "arXiv",
    primaryClass = "hep-ph",
    doi = "10.1142/S0217751X1530001X",
    journal = "Int. J. Mod. Phys. A",
    volume = "30",
    number = "04n05",
    pages = "1530001",
    year = "2015"
}

@article{Vergados:2016hso,
    author = "Vergados, J. D. and Ejiri, H. and {\v{S}}imkovic, F.",
    title = "{Neutrinoless double beta decay and neutrino mass}",
    eprint = "1612.02924",
    archivePrefix = "arXiv",
    primaryClass = "hep-ph",
    doi = "10.1142/S0218301316300071",
    journal = "Int. J. Mod. Phys. E",
    volume = "25",
    number = "11",
    pages = "1630007",
    year = "2016"
}

@article{Formaggio:2021nfz,
    author = "Formaggio, Joseph A. and de Gouv{\^e}a, Andr{\'e} Luiz C. and Robertson, R. G. Hamish",
    title = "{Direct Measurements of Neutrino Mass}",
    eprint = "2102.00594",
    archivePrefix = "arXiv",
    primaryClass = "nucl-ex",
    doi = "10.1016/j.physrep.2021.02.002",
    journal = "Phys. Rept.",
    volume = "914",
    pages = "1--54",
    year = "2021"
}

@article{KamLAND-Zen:2022tow,
    author = "Abe, S. and others",
    collaboration = "KamLAND-Zen",
    title = "{Search for the Majorana Nature of Neutrinos in the Inverted Mass Ordering Region with KamLAND-Zen}",
    eprint = "2203.02139",
    archivePrefix = "arXiv",
    primaryClass = "hep-ex",
    doi = "10.1103/PhysRevLett.130.051801",
    journal = "Phys. Rev. Lett.",
    volume = "130",
    number = "5",
    pages = "051801",
    year = "2023"
}

@article{nEXO:2017nam,
    author = "Albert, J. B. and others",
    collaboration = "nEXO",
    title = "{Sensitivity and Discovery Potential of nEXO to Neutrinoless Double Beta Decay}",
    eprint = "1710.05075",
    archivePrefix = "arXiv",
    primaryClass = "nucl-ex",
    reportNumber = "LLNL-JRNL-737682",
    doi = "10.1103/PhysRevC.97.065503",
    journal = "Phys. Rev. C",
    volume = "97",
    number = "6",
    pages = "065503",
    year = "2018"
}

@article{Meroni:2014tba,
    author = "Meroni, A. and Peinado, E.",
    title = "{The quest for neutrinoless double beta decay: Pseudo-Dirac, Majorana and sterile neutrinos}",
    eprint = "1406.3990",
    archivePrefix = "arXiv",
    primaryClass = "hep-ph",
    reportNumber = "RM3-TH-14-11",
    doi = "10.1103/PhysRevD.90.053002",
    journal = "Phys. Rev. D",
    volume = "90",
    number = "5",
    pages = "053002",
    year = "2014"
}

@article{Giganti:2017fhf,
    author = "Giganti, Claudio and Lavignac, St{\'e}phane and Zito, Marco",
    title = "{Neutrino oscillations: The rise of the PMNS paradigm}",
    eprint = "1710.00715",
    archivePrefix = "arXiv",
    primaryClass = "hep-ex",
    doi = "10.1016/j.ppnp.2017.10.001",
    journal = "Prog. Part. Nucl. Phys.",
    volume = "98",
    pages = "1--54",
    year = "2018"
}

@article{Qian:2015waa,
    author = "Qian, X. and Vogel, P.",
    title = "{Neutrino Mass Hierarchy}",
    eprint = "1505.01891",
    archivePrefix = "arXiv",
    primaryClass = "hep-ex",
    reportNumber = "BNL-108037-2015-JA",
    doi = "10.1016/j.ppnp.2015.05.002",
    journal = "Prog. Part. Nucl. Phys.",
    volume = "83",
    pages = "1--30",
    year = "2015"
}

@article{Capozzi:2025wyn,
    author = "Capozzi, Francesco and Giar{\`e}, William and Lisi, Eligio and Marrone, Antonio and Melchiorri, Alessandro and Palazzo, Antonio",
    title = "{Neutrino masses and mixing: Entering the era of subpercent precision}",
    eprint = "2503.07752",
    archivePrefix = "arXiv",
    primaryClass = "hep-ph",
    doi = "10.1103/PhysRevD.111.093006",
    journal = "Phys. Rev. D",
    volume = "111",
    number = "9",
    pages = "093006",
    year = "2025"
}

@article{Giunti:2015kza,
    author = "Giunti, C. and Zavanin, E. M.",
    title = "{Predictions for Neutrinoless Double-Beta Decay in the 3+1 Sterile Neutrino Scenario}",
    eprint = "1505.00978",
    archivePrefix = "arXiv",
    primaryClass = "hep-ph",
    doi = "10.1007/JHEP07(2015)171",
    journal = "JHEP",
    volume = "07",
    pages = "171",
    year = "2015"
}

@article{KATRIN:2024cdt,
    author = "Aker, Max and others",
    collaboration = "KATRIN",
    title = "{Direct neutrino-mass measurement based on 259 days of KATRIN data}",
    eprint = "2406.13516",
    archivePrefix = "arXiv",
    primaryClass = "nucl-ex",
    doi = "10.1126/science.adq9592",
    journal = "Science",
    volume = "388",
    number = "6743",
    pages = "adq9592",
    year = "2025"
}

@article{KATRIN:2021uub,
	author = "Aker, M. and others",
	collaboration = "KATRIN",
	title = "{Direct neutrino-mass measurement with sub-electronvolt sensitivity}",
	eprint = "2105.08533",
	archivePrefix = "arXiv",
	primaryClass = "hep-ex",
	doi = "10.1038/s41567-021-01463-1",
	journal = "Nature Phys.",
	volume = "18",
	number = "2",
	pages = "160--166",
	year = "2022"
}

@inproceedings{Project8:2022wqh,
    author = "Esfahani, A. Ashtari and others",
    collaboration = "Project 8",
    title = "{The Project 8 Neutrino Mass Experiment}",
    booktitle = "{Snowmass 2021}",
    eprint = "2203.07349",
    archivePrefix = "arXiv",
    primaryClass = "nucl-ex",
    month = "3",
    year = "2022"
}

@article{Kopp:2013vaa,
    author = "Kopp, Joachim and Machado, Pedro A. N. and Maltoni, Michele and Schwetz, Thomas",
    title = "{Sterile Neutrino Oscillations: The Global Picture}",
    eprint = "1303.3011",
    archivePrefix = "arXiv",
    primaryClass = "hep-ph",
    reportNumber = "IFT-UAM-CSIC-13-026",
    doi = "10.1007/JHEP05(2013)050",
    journal = "JHEP",
    volume = "05",
    pages = "050",
    year = "2013"
}

@article{Giunti:2013aea,
    author = "Giunti, C. and Laveder, M. and Li, Y. F. and Long, H. W.",
    title = "{Pragmatic View of Short-Baseline Neutrino Oscillations}",
    eprint = "1308.5288",
    archivePrefix = "arXiv",
    primaryClass = "hep-ph",
    doi = "10.1103/PhysRevD.88.073008",
    journal = "Phys. Rev. D",
    volume = "88",
    pages = "073008",
    year = "2013"
}

@article{KATRIN:2025lph,
	author = "Acharya, Himal and others",
	collaboration = "KATRIN",
	title = "{Sterile-neutrino search based on 259 days of KATRIN data}",
	eprint = "2503.18667",
	archivePrefix = "arXiv",
	primaryClass = "hep-ex",
	doi = "10.1038/s41586-025-09739-9",
	journal = "Nature",
	volume = "648",
	number = "8092",
	pages = "70--75",
	year = "2025"
}

@article{KamLAND-Zen:2024eml,
	author = "Abe, S. and others",
	collaboration = "KamLAND-Zen",
	title = "{Search for Majorana Neutrinos with the Complete KamLAND-Zen Dataset}",
	eprint = "2406.11438",
	archivePrefix = "arXiv",
	primaryClass = "hep-ex",
	doi = "10.1103/jkf6-48j8",
	journal = "Phys. Rev. Lett.",
	volume = "135",
	number = "26",
	pages = "262501",
	year = "2025"
}

@article{nEXO:2021ujk,
    author = "Adhikari, G. and others",
    collaboration = "nEXO",
    title = "{nEXO: neutrinoless double beta decay search beyond 10$^{28}$ year half-life sensitivity}",
    eprint = "2106.16243",
    archivePrefix = "arXiv",
    primaryClass = "nucl-ex",
    doi = "10.1088/1361-6471/ac3631",
    journal = "J. Phys. G",
    volume = "49",
    number = "1",
    pages = "015104",
    year = "2022"
}

@article{Brugnera:2023zgw,
    author = "Brugnera, Riccardo",
    collaboration = "LEGEND",
    title = "{Neutrinoless double beta decay search with LEGEND}",
    eprint = "2501.10046",
    archivePrefix = "arXiv",
    primaryClass = "nucl-ex",
    doi = "10.22323/1.421.0075",
    journal = "PoS",
    volume = "NOW2022",
    pages = "075",
    year = "2023"
}

@article{Gonzalez-Garcia:2002bkq,
    author = "Gonzalez-Garcia, M. C. and Nir, Yosef",
    title = "{Neutrino Masses and Mixing: Evidence and Implications}",
    eprint = "hep-ph/0202058",
    archivePrefix = "arXiv",
    reportNumber = "CERN-TH-2002-021, WIS-08-02-FEB-DPP",
    doi = "10.1103/RevModPhys.75.345",
    journal = "Rev. Mod. Phys.",
    volume = "75",
    pages = "345--402",
    year = "2003"
}

@article{SNO:2002tuh,
    author = "Ahmad, Q. R. and others",
    collaboration = "SNO",
    title = "{Direct evidence for neutrino flavor transformation from neutral current interactions in the Sudbury Neutrino Observatory}",
    eprint = "nucl-ex/0204008",
    archivePrefix = "arXiv",
    doi = "10.1103/PhysRevLett.89.011301",
    journal = "Phys. Rev. Lett.",
    volume = "89",
    pages = "011301",
    year = "2002"
}

@article{K2K:2002icj,
    author = "Ahn, M. H. and others",
    collaboration = "K2K",
    title = "{Indications of neutrino oscillation in a 250 km long baseline experiment}",
    eprint = "hep-ex/0212007",
    archivePrefix = "arXiv",
    doi = "10.1103/PhysRevLett.90.041801",
    journal = "Phys. Rev. Lett.",
    volume = "90",
    pages = "041801",
    year = "2003"
}

@article{KamLAND:2004mhv,
    author = "Araki, T. and others",
    collaboration = "KamLAND",
    title = "{Measurement of neutrino oscillation with KamLAND: Evidence of spectral distortion}",
    eprint = "hep-ex/0406035",
    archivePrefix = "arXiv",
    doi = "10.1103/PhysRevLett.94.081801",
    journal = "Phys. Rev. Lett.",
    volume = "94",
    pages = "081801",
    year = "2005"
}

@inproceedings{Jones:2021cga,
    author = "Jones, Benjamin J. P.",
    title = "{The Physics of Neutrinoless Double Beta Decay: A Primer}",
    eprint = "2108.09364",
    archivePrefix = "arXiv",
    primaryClass = "nucl-ex",
    month = "8",
    year = "2021"
}

@article{Huang:2023tjy,
    author = "Huang, Chao-Shang",
    title = "{Majorana CP violating phases}",
    eprint = "2302.01623",
    archivePrefix = "arXiv",
    primaryClass = "hep-ph",
    doi = "10.1016/j.nuclphysb.2024.116763",
    journal = "Nucl. Phys. B",
    volume = "1010",
    pages = "116763",
    year = "2025"
}

@article{Dery:2024lem,
    author = "Dery, Avital and Gori, Stefania and Grossman, Yuval and Ligeti, Zoltan",
    title = "{Majorana phases beyond neutrinoless double beta decay}",
    eprint = "2406.18647",
    archivePrefix = "arXiv",
    primaryClass = "hep-ph",
    doi = "10.1007/JHEP10(2024)100",
    journal = "JHEP",
    volume = "10",
    pages = "100",
    year = "2024"
}

@article{Benato:2015via,
    author = "Benato, Giovanni",
    title = "{Effective Majorana Mass and Neutrinoless Double Beta Decay}",
    eprint = "1510.01089",
    archivePrefix = "arXiv",
    primaryClass = "hep-ph",
    doi = "10.1140/epjc/s10052-015-3802-1",
    journal = "Eur. Phys. J. C",
    volume = "75",
    number = "11",
    pages = "563",
    year = "2015"
}

@article{Kotila:2012zza,
    author = "Kotila, J. and Iachello, F.",
    title = "{Phase space factors for double-$\beta$ decay}",
    eprint = "1209.5722",
    archivePrefix = "arXiv",
    primaryClass = "nucl-th",
    doi = "10.1103/PhysRevC.85.034316",
    journal = "Phys. Rev. C",
    volume = "85",
    pages = "034316",
    year = "2012"
}

@article{Mirea:2014dza,
    author = "Mirea, M. and Pahomi, T. and Stoica, S.",
    title = "{Phase Space Factors for Double Beta Decay: an up-date}",
    eprint = "1411.5506",
    archivePrefix = "arXiv",
    primaryClass = "nucl-th",
    month = "11",
    year = "2014"
}

@article{Agostini:2022zub,
    author = "Agostini, Matteo and Benato, Giovanni and Detwiler, Jason A. and Men{\'e}ndez, Javier and Vissani, Francesco",
    title = "{Toward the discovery of matter creation with neutrinoless {\ensuremath{\beta}}{\ensuremath{\beta}} decay}",
    eprint = "2202.01787",
    archivePrefix = "arXiv",
    primaryClass = "hep-ex",
    doi = "10.1103/RevModPhys.95.025002",
    journal = "Rev. Mod. Phys.",
    volume = "95",
    number = "2",
    pages = "025002",
    year = "2023"
}

@article{Faessler:2007hu,
    author = "Faessler, Amand and Fogli, G. L. and Lisi, E. and Rodin, V. and Rotunno, A. M. and Simkovic, F.",
    title = "{Overconstrained estimates of neutrinoless double beta decay within the QRPA}",
    eprint = "0711.3996",
    archivePrefix = "arXiv",
    primaryClass = "nucl-th",
    doi = "10.1088/0954-3899/35/7/075104",
    journal = "J. Phys. G",
    volume = "35",
    pages = "075104",
    year = "2008"
}

@article{Engel:2014pha,
    author = "Engel, J. and Simkovic, F. and Vogel, P.",
    title = "{Chiral Two-Body Currents and Neutrinoless Double-Beta Decay in the QRPA}",
    eprint = "1403.7860",
    archivePrefix = "arXiv",
    primaryClass = "nucl-th",
    doi = "10.1103/PhysRevC.89.064308",
    journal = "Phys. Rev. C",
    volume = "89",
    number = "6",
    pages = "064308",
    year = "2014"
}

@article{Kota:2025dzk,
    author = "Kota, V. K. B. and Sahu, R.",
    title = "{Statistical shell model for neutrinoless double {\ensuremath{\beta}}-decay nuclear transition matrix elements: Results for $^{76}$Ge, $^{82}$Se, $^{100}$Mo, $^{124}$Sn, $^{130}$Te and $^{136}$Xe}",
    eprint = "2507.02707",
    archivePrefix = "arXiv",
    primaryClass = "nucl-th",
    doi = "10.1142/S0218301325500569",
    journal = "Int. J. Mod. Phys. E",
    volume = "35",
    number = "01",
    pages = "2550056",
    year = "2026"
}

@article{Barea:2013bz,
    author = "Barea, J. and Kotila, J. and Iachello, F.",
    title = "{Nuclear matrix elements for double-$\beta$ decay}",
    eprint = "1301.4203",
    archivePrefix = "arXiv",
    primaryClass = "nucl-th",
    doi = "10.1103/PhysRevC.87.014315",
    journal = "Phys. Rev. C",
    volume = "87",
    number = "1",
    pages = "014315",
    year = "2013"
}

@article{Ding:2024obt,
    author = "Ding, Chen-rong and Li, Gang and Yao, Jiang-ming",
    title = "{Nuclear matrix elements of neutrinoless double-beta decay in covariant density functional theory with different mechanisms}",
    eprint = "2403.17722",
    archivePrefix = "arXiv",
    primaryClass = "nucl-th",
    doi = "10.1016/j.physletb.2024.138896",
    journal = "Phys. Lett. B",
    volume = "856",
    pages = "138896",
    year = "2024"
}

@article{Gariazzo:2019cym,
    author = "Gariazzo, Stefano",
    editor = "Nakahata, Masayuki",
    title = "{Light sterile neutrinos: the current picture from neutrino oscillations}",
    eprint = "1911.03463",
    archivePrefix = "arXiv",
    primaryClass = "hep-ph",
    doi = "10.1088/1742-6596/1468/1/012120",
    journal = "J. Phys. Conf. Ser.",
    volume = "1468",
    number = "1",
    pages = "012120",
    year = "2020"
}

@article{Gariazzo:2017fdh,
    author = "Gariazzo, S. and Giunti, C. and Laveder, M. and Li, Y. F.",
    title = "{Updated Global 3+1 Analysis of Short-BaseLine Neutrino Oscillations}",
    eprint = "1703.00860",
    archivePrefix = "arXiv",
    primaryClass = "hep-ph",
    doi = "10.1007/JHEP06(2017)135",
    journal = "JHEP",
    volume = "06",
    pages = "135",
    year = "2017"
}

@article{Giunti:2015iyr,
    author = "Giunti, C. and Zavanin, E. M.",
    title = "{Predictions for Neutrinoless Double-Beta Decay in the 3+1 Sterile Neutrino Scenario}",
    eprint = "1511.03838",
    archivePrefix = "arXiv",
    primaryClass = "hep-ph",
    doi = "10.1088/1742-6596/718/6/062074",
    journal = "J. Phys. Conf. Ser.",
    volume = "718",
    number = "6",
    pages = "062074",
    year = "2016"
}

@article{Huang:2019qvq,
    author = "Huang, Guo-Yuan and Zhou, Shun",
    title = "{Impact of an eV-mass sterile neutrino on the neutrinoless double-beta decays: A Bayesian analysis}",
    eprint = "1902.03839",
    archivePrefix = "arXiv",
    primaryClass = "hep-ph",
    doi = "10.1016/j.nuclphysb.2019.114691",
    journal = "Nucl. Phys. B",
    volume = "945",
    pages = "114691",
    year = "2019"
}

@article{Vissani:1999tu,
    author = "Vissani, Francesco",
    title = "{Signal of neutrinoless double beta decay, neutrino spectrum and oscillation scenarios}",
    eprint = "hep-ph/9906525",
    archivePrefix = "arXiv",
    doi = "10.1088/1126-6708/1999/06/022",
    journal = "JHEP",
    volume = "06",
    pages = "022",
    year = "1999"
}

@article{Kleesiek:2018mel,
    author = "Kleesiek, M. and others",
    title = "{$\beta$-Decay Spectrum, Response Function and Statistical Model for Neutrino Mass Measurements with the KATRIN Experiment}",
    eprint = "1806.00369",
    archivePrefix = "arXiv",
    primaryClass = "physics.data-an",
    doi = "10.1140/epjc/s10052-019-6686-7",
    journal = "Eur. Phys. J. C",
    volume = "79",
    number = "3",
    pages = "204",
    year = "2019"
}

@article{LSND:2001aii,
    author = "Aguilar, A. and others",
    collaboration = "LSND",
    title = "{Evidence for neutrino oscillations from the observation of $\bar{\nu}_e$ appearance in a $\bar{\nu}_\mu$
 beam}",
    eprint = "hep-ex/0104049",
    archivePrefix = "arXiv",
    doi = "10.1103/PhysRevD.64.112007",
    journal = "Phys. Rev. D",
    volume = "64",
    pages = "112007",
    year = "2001"
}

@article{MiniBooNE:2018esg,
    author = "Aguilar-Arevalo, A. A. and others",
    collaboration = "MiniBooNE",
    title = "{Significant Excess of ElectronLike Events in the MiniBooNE Short-Baseline Neutrino Experiment}",
    eprint = "1805.12028",
    archivePrefix = "arXiv",
    primaryClass = "hep-ex",
    reportNumber = "LA-UR-18-24586, FERMILAB-PUB-18-219-AD-PPD-ND",
    doi = "10.1103/PhysRevLett.121.221801",
    journal = "Phys. Rev. Lett.",
    volume = "121",
    number = "22",
    pages = "221801",
    year = "2018"
}

@article{Barinov:2021asz,
    author = "Barinov, V. V. and others",
    title = "{Results from the Baksan Experiment on Sterile Transitions (BEST)}",
    eprint = "2109.11482",
    archivePrefix = "arXiv",
    primaryClass = "nucl-ex",
    doi = "10.1103/PhysRevLett.128.232501",
    journal = "Phys. Rev. Lett.",
    volume = "128",
    number = "23",
    pages = "232501",
    year = "2022"
}

@article{Kaether:2010ag,
    author = "Kaether, F. and Hampel, W. and Heusser, G. and Kiko, J. and Kirsten, T.",
    title = "{Reanalysis of the GALLEX solar neutrino flux and source experiments}",
    eprint = "1001.2731",
    archivePrefix = "arXiv",
    primaryClass = "hep-ex",
    doi = "10.1016/j.physletb.2010.01.030",
    journal = "Phys. Lett. B",
    volume = "685",
    pages = "47--54",
    year = "2010"
}

@article{SAGE:2009eeu,
    author = "Abdurashitov, J. N. and others",
    collaboration = "SAGE",
    title = "{Measurement of the solar neutrino capture rate with gallium metal. III: Results for the 2002--2007 data-taking period}",
    eprint = "0901.2200",
    archivePrefix = "arXiv",
    primaryClass = "nucl-ex",
    doi = "10.1103/PhysRevC.80.015807",
    journal = "Phys. Rev. C",
    volume = "80",
    pages = "015807",
    year = "2009"
}

@article{Dentler:2018sju,
    author = "Dentler, Mona and Hern{\'a}ndez-Cabezudo, {\'A}lvaro and Kopp, Joachim and Machado, Pedro A. N. and Maltoni, Michele and Martinez-Soler, Ivan and Schwetz, Thomas",
    title = "{Updated Global Analysis of Neutrino Oscillations in the Presence of eV-Scale Sterile Neutrinos}",
    eprint = "1803.10661",
    archivePrefix = "arXiv",
    primaryClass = "hep-ph",
    reportNumber = "MITP-18-023, IFT-UAM-CSIC-18-033, MITP/18-023, FERMILAB-PUB-18-086-T, IFT-UAM/CSIC-18-033",
    doi = "10.1007/JHEP08(2018)010",
    journal = "JHEP",
    volume = "08",
    pages = "010",
    year = "2018"
}

@article{Goswami:2026qpl,
    author = "Goswami, Srubabati and M., Hemanth and Pachhar, Debashis and Rajeev, N.",
    title = "{Neutrino mass variables in 3 active and 2 sterile neutrino scenario}",
    eprint = "2603.17689",
    archivePrefix = "arXiv",
    primaryClass = "hep-ph",
    month = "3",
    year = "2026"
}

@article{Jana:2024xmc,
    author = "Jana, Sudip and Puetter, Lucas and Smirnov, Alexei Yu.",
    title = "{Restricting sterile neutrinos by neutrinoless double beta decay}",
    eprint = "2408.01488",
    archivePrefix = "arXiv",
    primaryClass = "hep-ph",
    doi = "10.1103/PhysRevD.111.015011",
    journal = "Phys. Rev. D",
    volume = "111",
    number = "1",
    pages = "015011",
    year = "2025"
}

@article{Goswami:2005ng,
    author = "Goswami, Srubabati and Rodejohann, Werner",
    title = "{Constraining mass spectra with sterile neutrinos from neutrinoless double beta decay, tritium beta decay and cosmology}",
    eprint = "hep-ph/0512234",
    archivePrefix = "arXiv",
    reportNumber = "TUM-HEP-615-05",
    doi = "10.1103/PhysRevD.73.113003",
    journal = "Phys. Rev. D",
    volume = "73",
    pages = "113003",
    year = "2006"
}

@article{Farzan:2002zq,
    author = "Farzan, Y. and Smirnov, A. Yu.",
    title = "{On the effective mass of the electron neutrino in beta decay}",
    eprint = "hep-ph/0211341",
    archivePrefix = "arXiv",
    reportNumber = "SLAC-PUB-9593, SISSA-70-2002-EP",
    doi = "10.1016/S0370-2693(03)00207-7",
    journal = "Phys. Lett. B",
    volume = "557",
    pages = "224--232",
    year = "2003"
}

@article{Chambyal:2026fnq,
    author = "Chambyal, Rushi and Tapender and Singh, Labh and Verma, Surender",
    title = "{Phenomenology of Vanishing Effective Majorana Mass with a Sterile Neutrino under Cosmological and JUNO Constraints}",
    eprint = "2604.14983",
    archivePrefix = "arXiv",
    primaryClass = "hep-ph",
    month = "4",
    year = "2026"
}

@article{Giunti:2019fcj,
    author = "Giunti, C. and Li, Y. F. and Zhang, Y. Y.",
    title = "{KATRIN bound on 3+1 active-sterile neutrino mixing and the reactor antineutrino anomaly}",
    eprint = "1912.12956",
    archivePrefix = "arXiv",
    primaryClass = "hep-ph",
    doi = "10.1007/JHEP05(2020)061",
    journal = "JHEP",
    volume = "05",
    pages = "061",
    year = "2020"
}

@article{Elbers:2025vlz,
    author = "Elbers, W. and others",
    title = "{Constraints on neutrino physics from DESI DR2 BAO and DR1 full shape}",
    eprint = "2503.14744",
    archivePrefix = "arXiv",
    primaryClass = "astro-ph.CO",
    doi = "10.1103/w9pk-xsk7",
    journal = "Phys. Rev. D",
    volume = "112",
    number = "8",
    pages = "083513",
    year = "2025"
}

@article{Naredo-Tuero:2024sgf,
    author = "Naredo-Tuero, Daniel and Escudero, Miguel and Fern{\'a}ndez-Mart{\'\i}nez, Enrique and Marcano, Xabier and Poulin, Vivian",
    title = "{Critical look at the cosmological neutrino mass bound}",
    eprint = "2407.13831",
    archivePrefix = "arXiv",
    primaryClass = "astro-ph.CO",
    doi = "10.1103/PhysRevD.110.123537",
    journal = "Phys. Rev. D",
    volume = "110",
    number = "12",
    pages = "123537",
    year = "2024"
}

@article{SPT-3G:2025bzu,
    author = "Camphuis, E. and others",
    collaboration = "SPT-3G",
    title = "{SPT-3G D1: CMB temperature and polarization power spectra and cosmology from 2019 and 2020 observations of the SPT-3G main field}",
    eprint = "2506.20707",
    archivePrefix = "arXiv",
    primaryClass = "astro-ph.CO",
    doi = "10.1103/7wt3-9v2y",
    journal = "Phys. Rev. D",
    volume = "113",
    number = "8",
    pages = "083504",
    year = "2026"
}

@article{JUNO:2025gmd,
    author = "Abusleme, Angel and others",
    collaboration = "JUNO",
    title = "{Measurement of reactor neutrino oscillation with the first JUNO data}",
    eprint = "2511.14593",
    archivePrefix = "arXiv",
    primaryClass = "hep-ex",
    doi = "10.1038/s41586-026-10538-z",
    journal = "Nature",
    volume = "654",
    number = "8118",
    pages = "343--348",
    year = "2026"
}

@article{Ge:2025cky,
    author = "Ge, Shao-Feng and Kong, Chui-Fan and Lindner, Manfred and Pinheiro, Jo{\~a}o Paulo",
    title = "{Neutrinoless double beta decay in light of JUNO first data}",
    eprint = "2511.15391",
    archivePrefix = "arXiv",
    primaryClass = "hep-ph",
    doi = "10.1007/JHEP03(2026)105",
    journal = "JHEP",
    volume = "03",
    pages = "105",
    year = "2026"
}

@article{CUPID:2025faj,
    author = "Alfonso, K. and others",
    collaboration = "CUPID",
    title = "{Sensitivity of the CUPID experiment to $0\nu \beta \beta $ decay of $^{100}$Mo}",
    eprint = "2504.14369",
    archivePrefix = "arXiv",
    primaryClass = "nucl-ex",
    doi = "10.1140/epjc/s10052-026-15296-w",
    journal = "Eur. Phys. J. C",
    volume = "86",
    number = "6",
    pages = "633",
    year = "2026"
}

@article{Capozzi:2025ovi,
    author = "Capozzi, Francesco and Lisi, Eligio and Marcone, Francesco and Marrone, Antonio and Palazzo, Antonio",
    title = "{Updated bounds on the (1, 2) neutrino oscillation parameters after first JUNO results}",
    eprint = "2511.21650",
    archivePrefix = "arXiv",
    primaryClass = "hep-ph",
    doi = "10.1103/cxqw-1bty",
    journal = "Phys. Rev. D",
    volume = "114",
    number = "1",
    pages = "016026",
    year = "2026"
}

@article{JUNO:2022mxj,
	author = "Abusleme, Angel and others",
	collaboration = "JUNO",
	title = "{Sub-percent precision measurement of neutrino oscillation parameters with JUNO}",
	eprint = "2204.13249",
	archivePrefix = "arXiv",
	primaryClass = "hep-ex",
	doi = "10.1088/1674-1137/ac8bc9",
	journal = "Chin. Phys. C",
	volume = "46",
	number = "12",
	pages = "123001",
	year = "2022"
}

@article{CUORE:2024ikf,
	author = "Adams, D. Q. and others",
	collaboration = "CUORE",
	title = "{Constraints on lepton number violation with the 2 tonne {\textperiodcentered} year CUORE dataset}",
	eprint = "2404.04453",
	archivePrefix = "arXiv",
	primaryClass = "nucl-ex",
	doi = "10.1126/science.adp6474",
	journal = "Science",
	volume = "390",
	number = "6777",
	pages = "1029--1032",
	year = "2025"
}

\end{document}